\RequirePackage{fix-cm}
\documentclass[smallextended]{svjour3}       
\smartqed  
\usepackage{graphicx}
\usepackage{natbib}
\usepackage{hyperref}
\usepackage{amsmath}
\usepackage{aas_macros}
\usepackage{mathabx}
\usepackage{multirow}
\journalname{Experimental Astronomy}
\begin{document}

\title{Background for a gamma-ray satellite on a low-Earth orbit
}


\author{P. Cumani         \and
        M. Hernanz       \and
        J. Kiener        \and
        V. Tatischeff   \and
        A. Zoglauer     
}


\institute{
        P. Cumani \at
              CSNSM, CNRS and University of Paris Sud, F-91405, Orsay, France \\
              \email{paolo.cumani@csnsm.in2p3.fr}         \and
        M. Hernanz   \at   
        Institute of Space Sciences (ICE, CSIC) and IEEC, Campus UAB, Cam\'i de Can Magrans s/n, 08193 Cerdanyola del Vall\`es (Barcelona), Spain \and
        J. Kiener,  V. Tatischeff\at
              CSNSM, CNRS and University of Paris Sud, F-91405, Orsay, France \\   \and
        A. Zoglauer     \at
        University of California at Berkeley, Space Sciences Laboratory, 7 Gauss Way, Berkeley, CA 94720, USA
}

\date{Received: date / Accepted: date}

\maketitle

\begin{abstract}
The different background components in a low-Earth orbit have been modeled in the 10 keV to 100 GeV energy range. The model is based on data from previous instruments and it considers both primary and secondary particles, charged particles, neutrons and photons. The necessary corrections to consider the geomagnetic cutoff are applied to calculate the flux at different inclinations and altitudes for a mean solar activity. Activation simulations from such a background have been carried out using the model of a possible future gamma-ray mission (e-ASTROGAM). The event rates and spectra from these simulations were then compared to those from the isotopes created by the particles present in the South Atlantic Anomaly (SAA). The primary protons are found to be the main contributor of the activation, while the contributions of the neutrons, and that of the secondary protons can be considered negligible. The long-term activation from the passage through the SAA becomes the main source of background at high inclination (i$\gtrsim10^\circ$). The used models have been collected in a Python class openly available on github.
\end{abstract}

\section{Introduction}\label{s:intro}

Conventionally, Low-Earth Orbits (LEOs) have a maximum altitude of $\sim 2000$ km, with a $\sim 300$ km lower limit given by the rapid orbital decay caused by the atmospheric drag. The inner Van Allen radiation belt (\cite{allen1958observation}), with a high level of trapped radiation, extends beyond $\sim 1000$ km. Satellites positioned at lower altitudes benefit, at low geomagnetic latitudes, of the shielding of the Earth's magnetic field, blocking a large part of the charged particles coming from the interplanetary medium.\\
As a drawback, all LEOs are subject to the drag caused by the remaining atmosphere and to the background from albedo and secondary particles created in the atmosphere.\\

Focusing on the characteristics of a gamma-ray mission in a low inclination orbit, a model for each component of the background in the 10 keV to 100~GeV energy range will be described in section \ref{s:model}. The South Atlantic Anomaly (SAA), the area in which the inner Van Allen belt comes closer to Earth, will be illustrated in section \ref{s:SAA}, simulating the particle spectrum along a typical orbit at an altitude of 550 km. The computed spectra will be used in section \ref{s:AcSim} to simulate the activation of a possible future gamma-ray satellite, e-ASTROGAM, in order to estimate the contribution of the different components to the overall background and the importance of the radioisotopes created during the passage of the SAA for different inclinations. Conclusions are given in section \ref{s:Con}.

\section{Background models}\label{s:model}
\begin{figure}[ht]
\begin{center}
\includegraphics[width=0.95\textwidth]{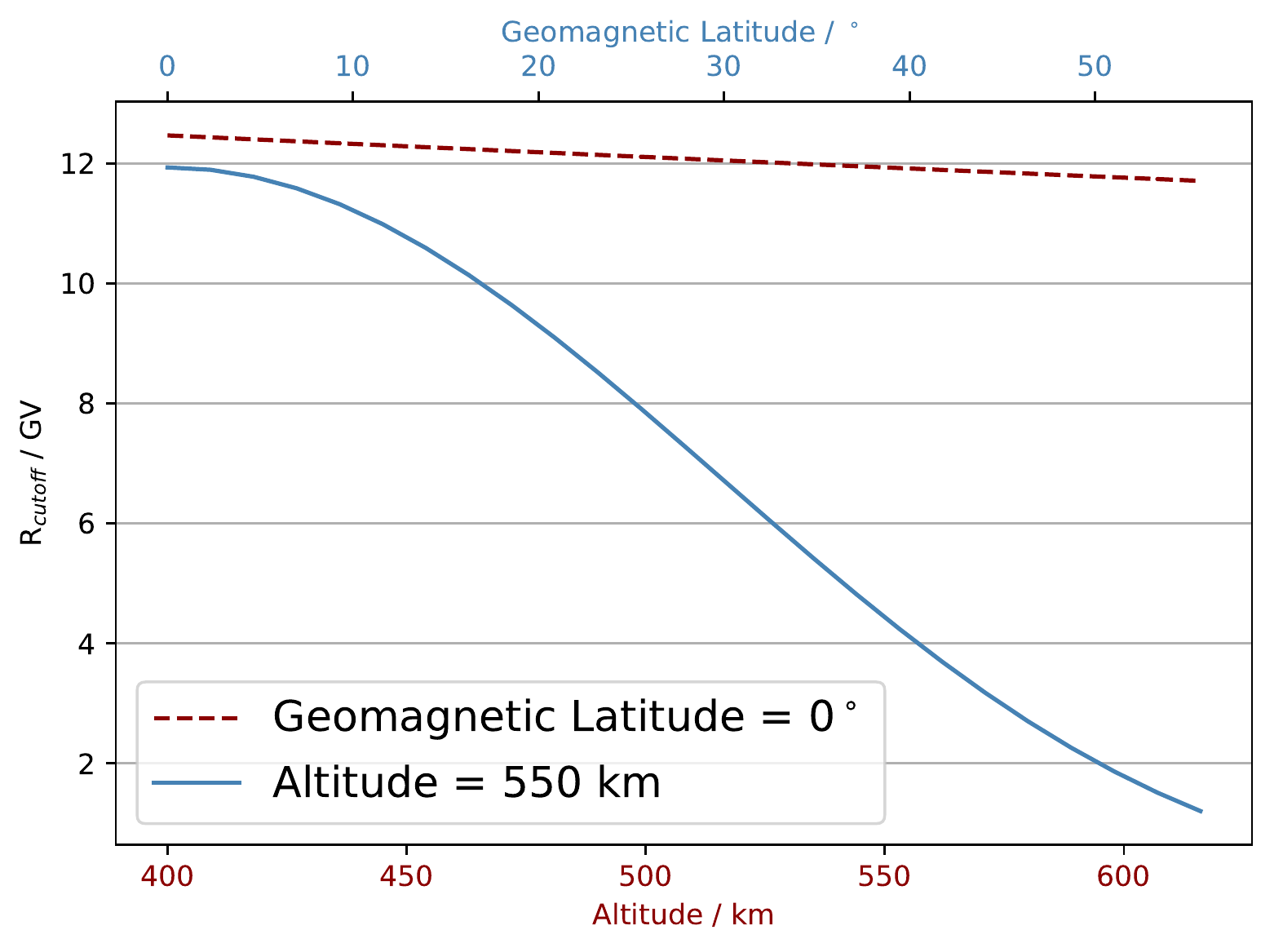}
\caption{Variation of the average geomagnetic cutoff $R_{cutoff}$ with altitude for constant latitude (0$^\circ$, red dashed line), and latitude for constant altitude (550~km, blue straight line).}
\label{fig:rcutoff}
\end{center}
\end{figure}

The LEO background spectrum depends mainly on the orbit altitude and inclination as well as the current solar activity.\\
In the following, where not differently specified, an altitude $h = 550$ km, inclination of the orbit $i = 0^\circ$, and a solar modulation potential value of 650~MV, corresponding to an activity halfway between minimum and maximum, will be used. The magnetic latitude is approximated to be equal, on average, to the inclination of the orbit.\\
Some of the parameters that are used to calculate more than one component are computed in the following way:
\begin{itemize}
    \item Horizon angle (in degree): defined as the angle between the zenith and the top of the atmosphere ($H_A=40$ km from sea level) at the satellite altitude $h$:
    \begin{equation}
        \theta _H = 90^\circ + \arccos\frac{R_\Earth + H_A}{R_\Earth+h}
    \end{equation}
    where $R_\Earth$ is the Earth's radius. The horizon angle for a 550 km equatorial orbit is equal to $112^\circ$.
    \item Average geomagnetic cutoff (in GV): is calculated as in \cite{SMART20052012}:
    \begin{equation}\label{eq:RcutFull}
        R_{cutoff} = \frac{M_d \cdot \cos ^4 \lambda}{r^2}\cdot \frac{1}{(1+\sqrt{1-\sin \epsilon \sin \xi \cos ^3 \lambda})^2}
    \end{equation}
    where $M_d$ is the magnitude of Earth's dipole moment, $\lambda$ is the latitude from the magnetic equator, $\epsilon$ is the angle from
the zenith direction, $\xi$ is the azimuthal angle measured clockwise from the direction to the north magnetic pole and $r$ is the distance from the dipole center. $r$ can be expressed as $R_\Earth+h$. Since we are computing the omnidirectional flux, $\xi$ can be set to zero. $M_d$ can be calculated by multiplying the $g^0 _1$ term of the International Geomagnetic Reference Field (IGRF) by the cube of the average radius of the Earth. The value from the IGRF-12\footnote{\url{https://www.ngdc.noaa.gov/IAGA/vmod/igrf.html}} for the year 2015 ($g^0 _1 = 29442.0$ nT) has been used\footnote{It must be noted that this value changes slowly, 0.5\% in the last 10 years, but constantly due to the evolution of Earth's magnetic field}. Equation \ref{eq:RcutFull} can then be simplified to:
    \begin{equation}\label{eq:Rcut}
    \begin{split}
    R_{cutoff} &= \frac{g^0 _1 \cdot R_\Earth }{4}\cdot \left(1+\frac{h}{R_\Earth}\right)^{-2}\cos ^4 \lambda \\
    &= 11.9 ~\mathrm{GV} \cdot \frac{g^0 _1}{29442.0 ~\mathrm{nT}} \cdot \left(\frac{R_\Earth+550 ~\mathrm{km}}{R_\Earth+h}\right)^2 \cos ^4 \lambda
    \end{split}
    \end{equation}

As shown in Fig. \ref{fig:rcutoff}, the average geomagnetic cutoff varies more steeply with geomagnetic latitude than altitude. 
\end{itemize}
In the following, the results of all the flux equations will be expressed as counts keV$^{-1}$ cm$^{-2}$ s$^{-1}$ sr$^{-1}$.
All the models of the different components described in the following have been collected in a Python class in an open github repository\footnote{\url{https://github.com/pcumani/LEOBackground}}.

\subsection{Extragalactic Photons}\label{s:cosmicphotons}
\begin{figure}[ht]
\begin{center}
\includegraphics[width=0.95\textwidth]{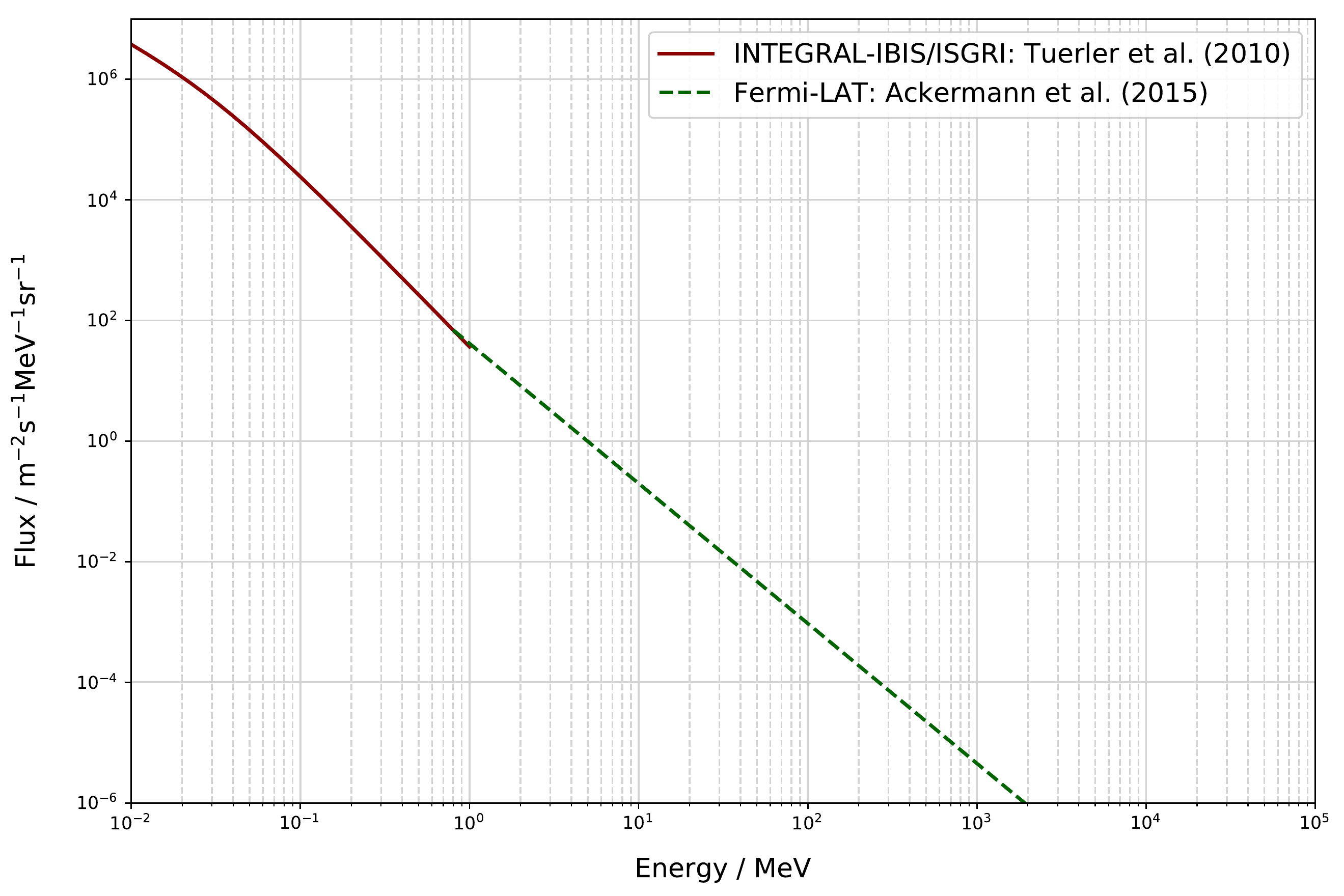}
\caption{Total extragalactic photons flux as modeled by INTEGRAL-IBIS/ISGRI, at low energies, and FERMI-LAT, at high energy.}
\label{fig:cosmicgamma}
\end{center}
\end{figure}

The isotropic extragalactic X and gamma-ray emission is both a significant component of the background for observation far from the Galactic plane, and an important science topic, especially in the MeV energy range (see, e.g., \cite{2017arXiv171101265D}).\\
The spectra shown in Fig. \ref{fig:cosmicgamma} are obtained using:
\begin{itemize}
    \item $E<890$ keV: From an extrapolation of the equation in \cite{2010A&A...512A..49T}, describing data of the IBIS/ISGRI gamma-ray imager instrument of INTEGRAL (\cite{2003A&A...411L...1W}): \begin{equation}\label{eq:turlerCP}
        F = \frac{0.109}{(E/28 ~\mathrm{keV})^{1.4}+(E/28 ~\mathrm{keV})^{2.88}}
    \end{equation}
    \item $E\geqslant890$ keV: Using the Fermi-LAT (\cite{2009ApJ...697.1071A}) results of the fit for the so-called foreground model A from \cite{2015ApJ...799...86A} and extrapolating it to lower energies from $E\geqslant100$ MeV:
    \begin{equation}\label{eq:AckermannCP}
    \begin{split}
        F&=I_{100} \left(\frac{E}{100~\mathrm{MeV}}\right)^{-\gamma} \exp{\frac{-E}{E_{cut}}}\\
        &=0.95\cdot 10^{-10} \left(\frac{E}{100~\mathrm{MeV}}\right)^{-2.32}\exp{\left(\frac{-E}{279\cdot 10^3~\mathrm{MeV}}\right)}
    \end{split}
    \end{equation}
    The extrapolated Fermi-LAT data are in good agreement with EGRET data (\cite{0004-637X-494-2-523}, see equation 18 from \cite{0004-637X-614-2-1113}).
\end{itemize}

\subsection{Galactic plane}

\begin{figure}[ht]
\begin{center}
\includegraphics[width=0.95\textwidth]{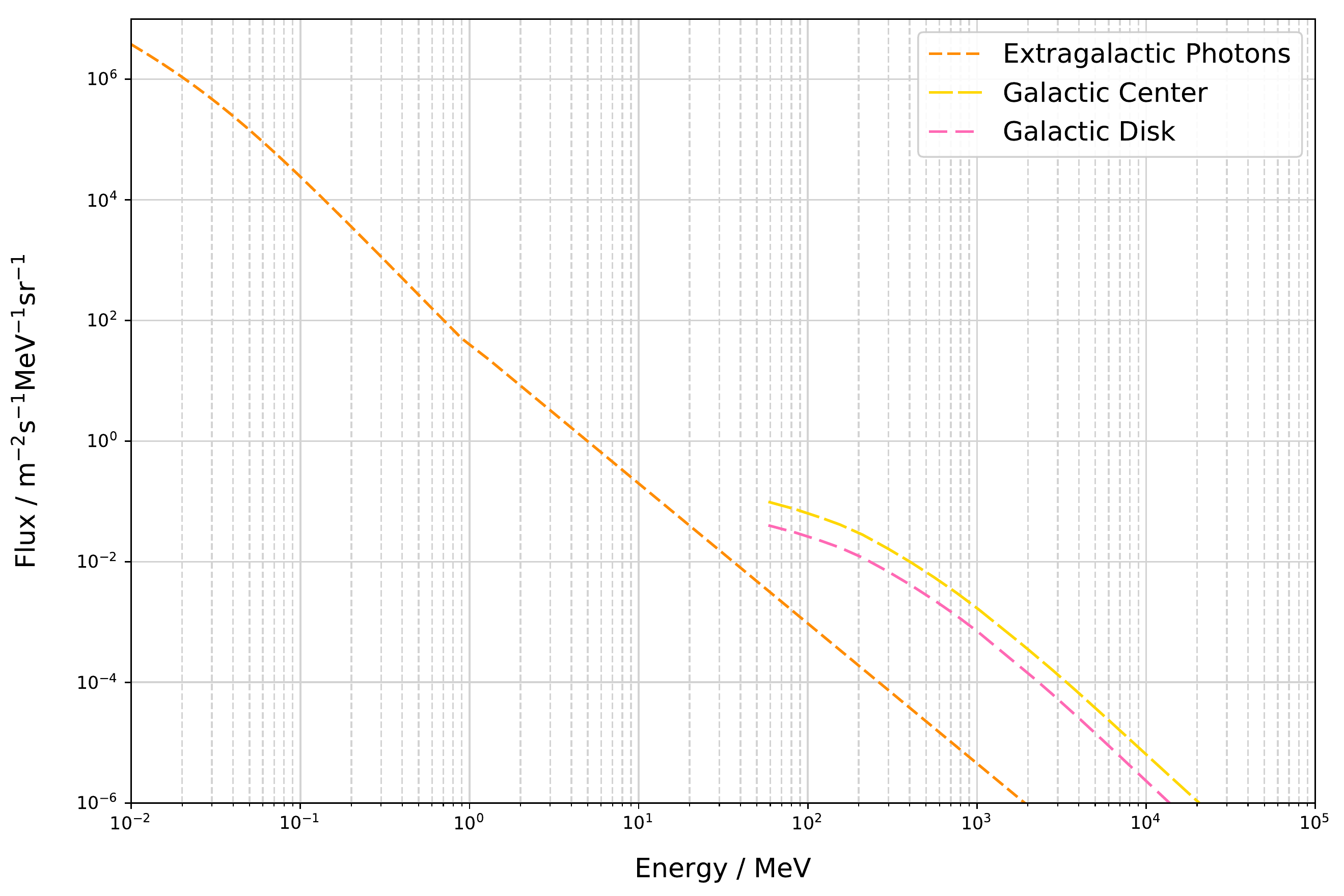}
\includegraphics[width=0.95\textwidth]{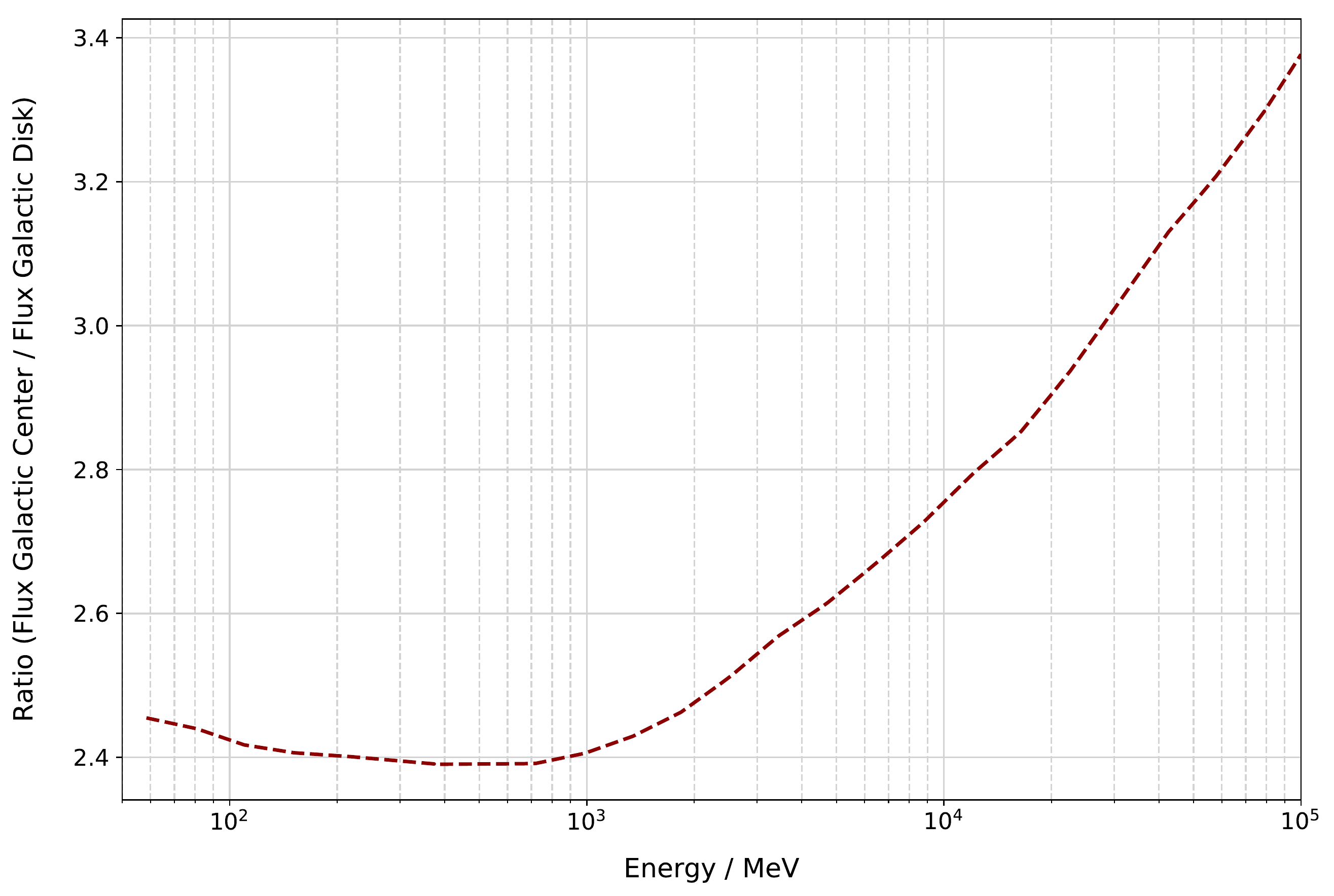}
\caption{Extragalactic photon spectrum compared with the background spectrum from the Galactic center and the Galactic disk (\textit{top}), and the ratio between these last two (\textit{bottom}).}
\label{fig:galactic}
\end{center}
\end{figure}

\begin{figure}[ht]
\begin{center}
\includegraphics[width=\textwidth]{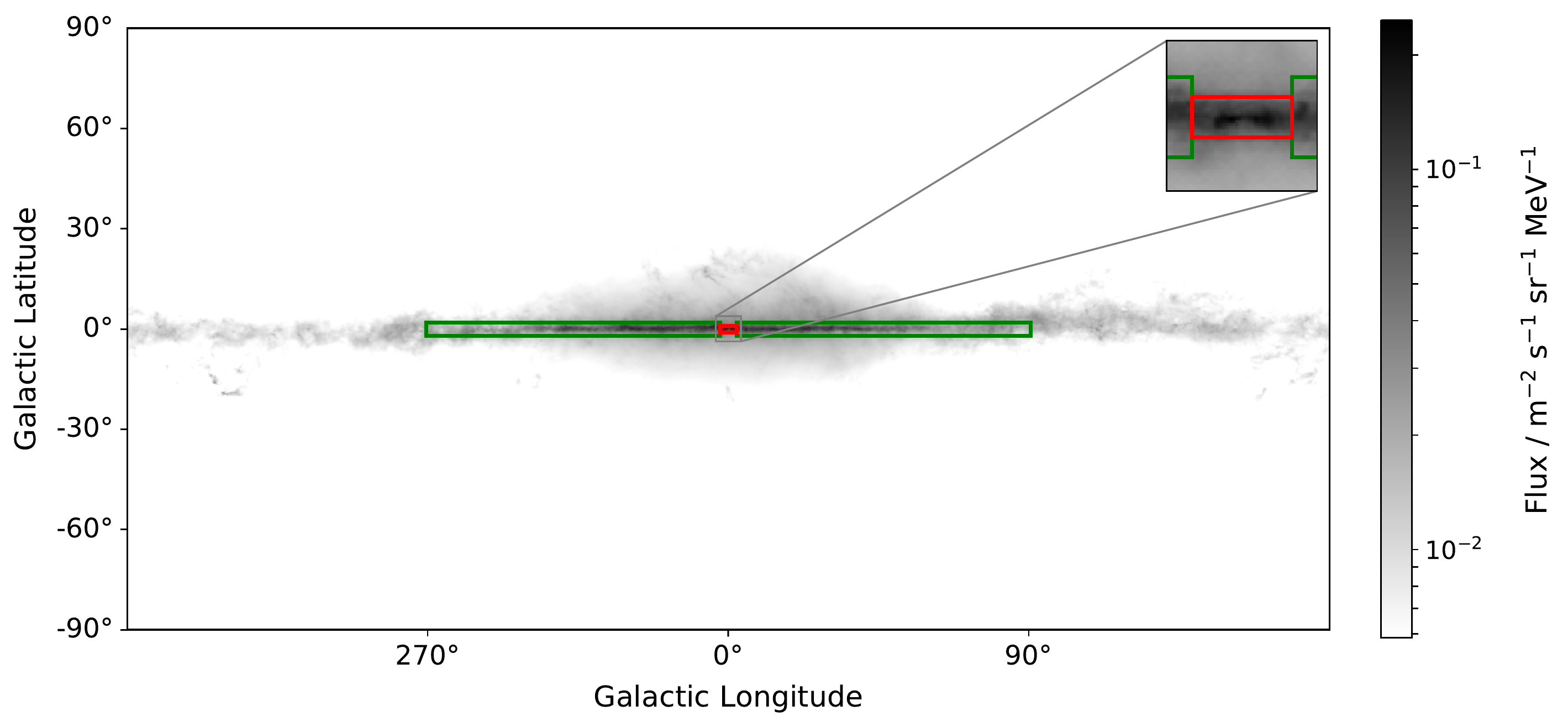}
\caption{The two regions defined as Galactic center (in red) and Galactic disk (in green) in the text, superimposed on the publicly available Fermi-LAT background at 58 MeV. The inset shows a zoom on the Galactic center region}
\label{fig:galaxyregion}
\end{center}
\end{figure}

It's useful, especially at high energy, to study separately the gamma-ray background for observations of the Galactic plane region. Such background extends up to the highest considered energies and, starting from $\sim200$ MeV, is more than an order of magnitude larger than the isotropic background previously described.\\
It's worth noting that the signal-to-noise ratio is directly related to the performance of the instruments. Thanks to a higher angular resolution, new instruments could be able to discriminate several point sources in previously confused areas, either Galactic or not.\\
Currently, the most sensitive gamma-ray experiment in the considered energy range is the Fermi-LAT. The Pass 8 interstellar emission model\footnote{\url{https://fermi.gsfc.nasa.gov/ssc/data/access/lat/BackgroundModels.html}} was used to obtain the results shown in Fig \ref{fig:galactic}. For each one of the 30 energies provided, starting from $\sim58$ MeV, two regions has been defined as shown in Fig. \ref{fig:galaxyregion}:
\begin{itemize}
    \item Galactic center: the most inner part of the Galaxy, defined by the coordinate $-1^\circ < b <1^\circ$, $-2.5^\circ < l < 2.5^\circ$.
    \item Galactic disk: a region defined as $-2^\circ < b <2^\circ$, $-90^\circ < l < 90^\circ$, excluding the $-2.5^\circ < l < 2.5^\circ$ area around the Galactic center.
\end{itemize}
The average flux in these two regions was computed at each energy. The ratio between the two regions varies between a factor of 2.4 to 3.4, increasing with energy, as shown at the bottom of Fig. \ref{fig:galactic}.

\subsection{Albedo Photons}
\begin{figure}[ht]
\begin{center}
\includegraphics[width=0.95\textwidth]{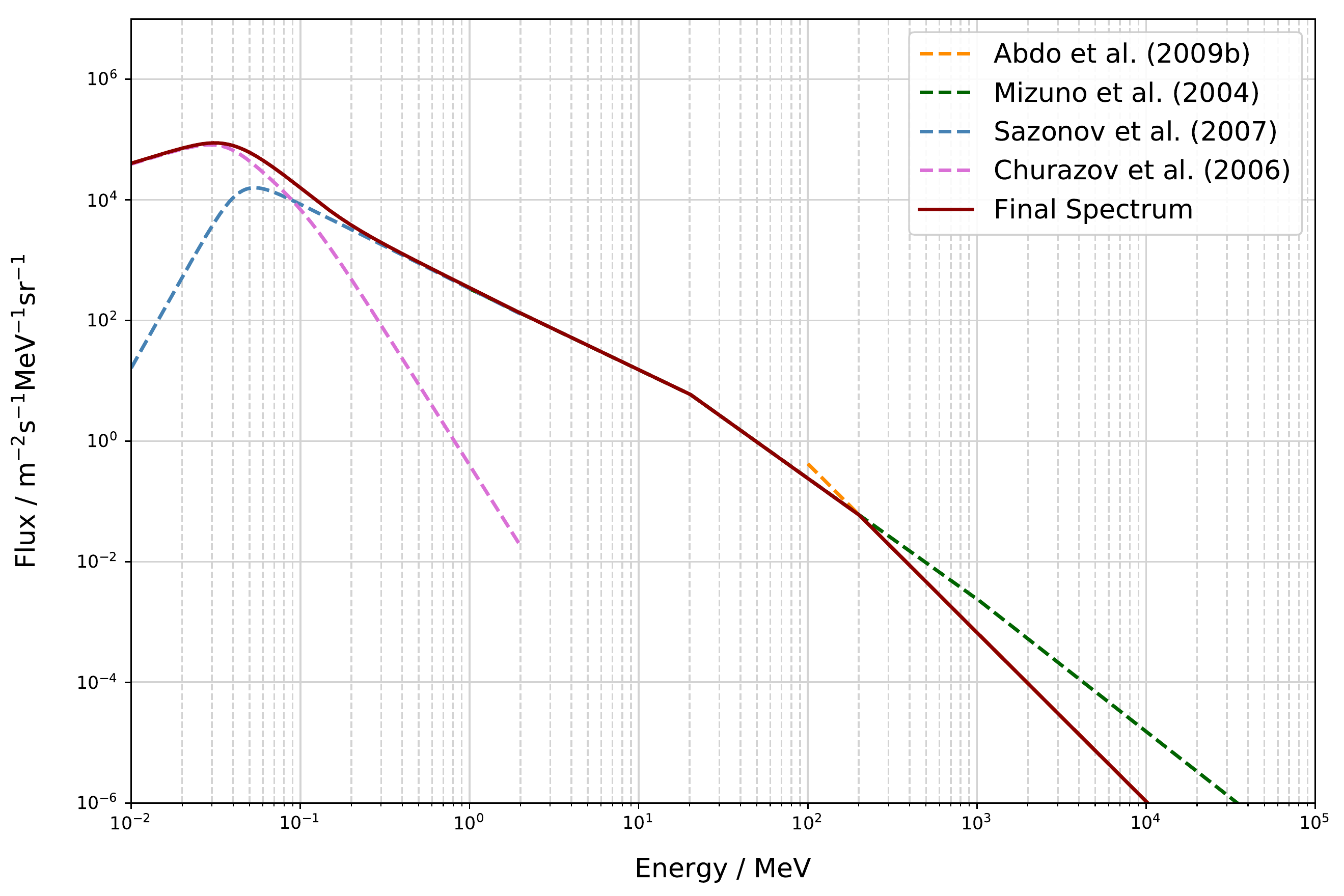}
\caption{Total albedo photons spectrum divided by the solid angle of the atmosphere ($\sim 4$~sr at 550 km). The different models composing it are also showed (see text).}
\label{fig:photons}
\end{center}
\end{figure}

Albedo photons are created by the Earth atmosphere, as a result of the interaction of cosmic rays or the reflection of the cosmic X-ray background. 
To model the zenith dependence of the spectrum, following the approach used by \cite{0004-637X-614-2-1113}, an angular distribution peaking at the top of the atmosphere ($\sim112^\circ$ at 550 km), to reproduce the limb-brightening observed in experimental data (see, e.g., \cite{2014PhRvL.112o1103A}), was used. The flux gradually decreases at higher angles reaching 50\% of its peak value at $\sim130^\circ$ (for an altitude of 550 km) and a minimum, one third of the maximum, at the nadir.\\
The results shown in Fig. \ref{fig:photons} are obtained using:
\begin{itemize}
    \item $E<1.85$ MeV: The sum of two components, both used in the analysis of INTEGRAL observations of the Earth, resulting in a good agreement between data and prediction: 
    \begin{itemize}
        \item[\textbullet] To describe the hard X-ray surface brightness of the Earth's atmosphere we used the output of equation 1 from \cite{2007MNRAS.377.1726S}:
        \begin{equation}
            F = \frac{C}{(E/44~\mathrm{keV})^{-5}+(E/44~\mathrm{keV})^{1.4}}
        \end{equation}
        where $C$ is:
        \begin{equation}
            C = \frac{3\mu(1+\mu)}{5\pi}\cdot \frac{1.47\cdot0.0178[(\phi/2.8)^{0.4}+(\phi/2.8)^{1.5}]^{-1}}{\sqrt{1+\{R_{cutoff}/[1.3\phi^{0.25}(1+2.5\phi^{0.4})]\}^2}}
        \end{equation}
        where $\mu = cos(180-\theta_H)$ is the cosine of the zenith angle and $\phi$ is the solar modulation potential in GV.
        \item[\textbullet] the reflected  X-ray background $F=F_0\cdot\Omega \cdot A(E)$, where $A(E)$ is calculated with equation 9 in \cite{2006astro.ph..8252C}:
        \begin{equation}
            \begin{split}
                A(E) &= \frac{1.22}{\left(\frac{E}{28.5}\right)^{-2.54}+\left(\frac{E}{51.3}\right)^{1.57}-0.37}\\
                & \cdot \frac{2.93+\left(\frac{E}{3.08}\right)^{4}}{1+\left(\frac{E}{3.08}\right)^{4}}\\
                & \cdot \frac{0.123+\left(\frac{E}{91.83}\right)^{3.44}}{1+\left(\frac{E}{91.83}\right)^{3.44}}
            \end{split}
        \end{equation}
        being E the energy in keV, and $F_0$ with equation \ref{eq:turlerCP} from section \ref{s:cosmicphotons}.
    \end{itemize}
    \item 1.85 MeV $< E <$ 200 MeV: from SAS 2 data (\cite{1981JGR....86.1265T}) as described in \cite{0004-637X-614-2-1113}:
    \begin{itemize}
\item[\textbullet] $E < 20$ MeV: \begin{equation}\label{eq:sas2-1}
    F= \frac{1010}{10^7}\cdot\left(\frac{E}{1 \mathrm{MeV}} \right)^{-1.34}\cdot\left(\frac{R_{cutoff}}{4.5~\mathrm{GV}}\right)^{-1.13}
\end{equation}
\item[\textbullet] 20 MeV $< E <$ 200 MeV: \begin{equation}\label{eq:sas2-2}
    F= \frac{7290}{10^7}\cdot\left(\frac{E}{1 ~\mathrm{MeV}} \right)^{-2}\cdot\left(\frac{R_{cutoff}}{4.5~\mathrm{GV}}\right)^{-1.13}
    \end{equation}
\end{itemize}
    \item $E>200$ MeV: The high energy component is taken from Fermi-LAT data (\cite{2009PhRvD..80l2004A}):
    \begin{equation}
    F= 1.823\cdot 10^{-8}\cdot\left(\frac{E}{200 \mathrm{MeV}} \right)^{-2.8}
    \end{equation}
\end{itemize}
The INTEGRAL and Fermi-LAT data are normalized to the results of equations \ref{eq:sas2-1} and \ref{eq:sas2-2} at, respectively, 1.85 and 200 MeV, to take into account the effect of the geomagnetic cutoff.

\subsection{Primary Protons}
\begin{figure}[ht]
\begin{center}
\includegraphics[width=\textwidth]{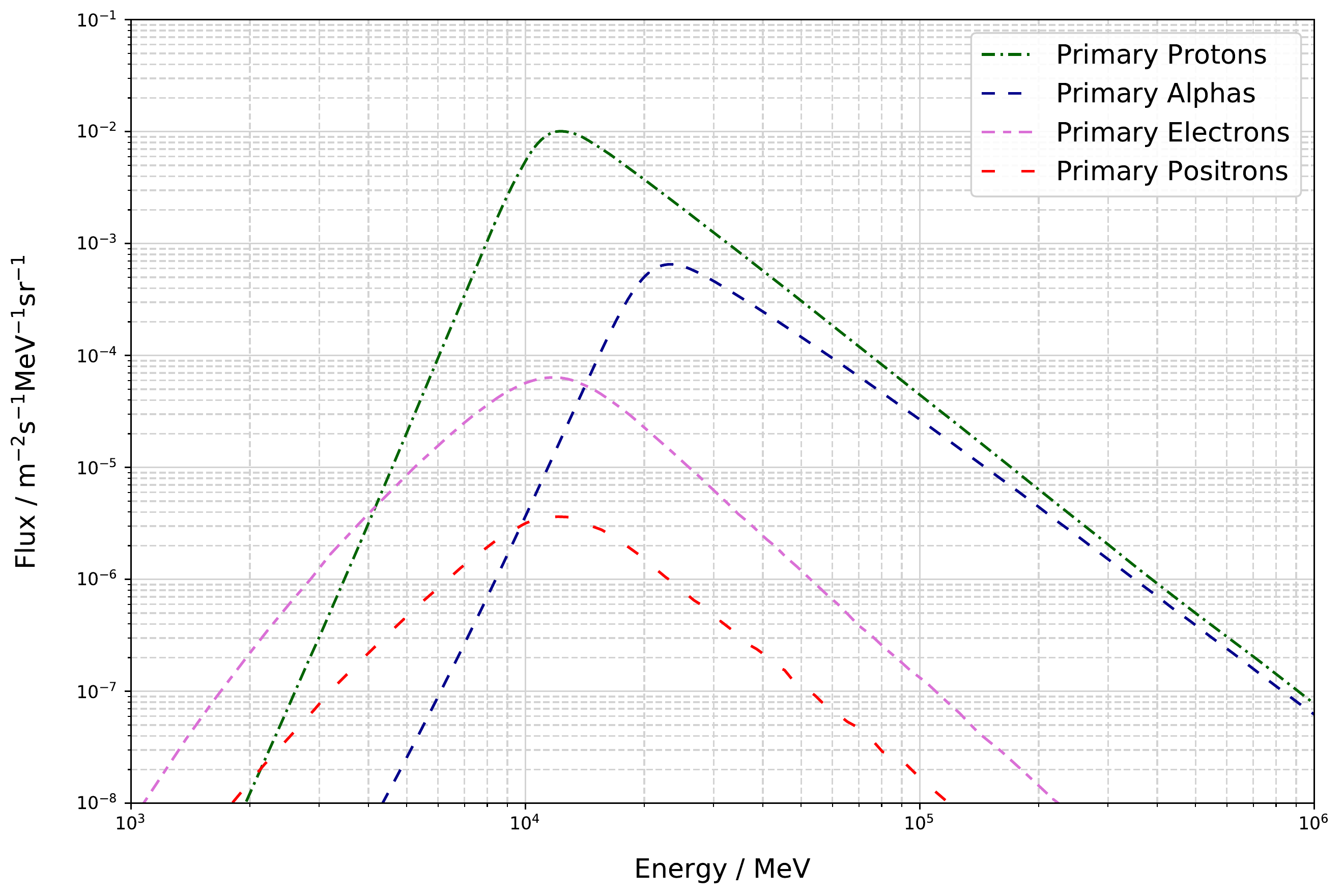}
\caption{Spectrum of the primary charged particles: protons, alphas, electrons, and positrons.}
\label{fig:prcharged}
\end{center}
\end{figure}
\begin{figure}[ht]
\begin{center}
\includegraphics[width=\textwidth]{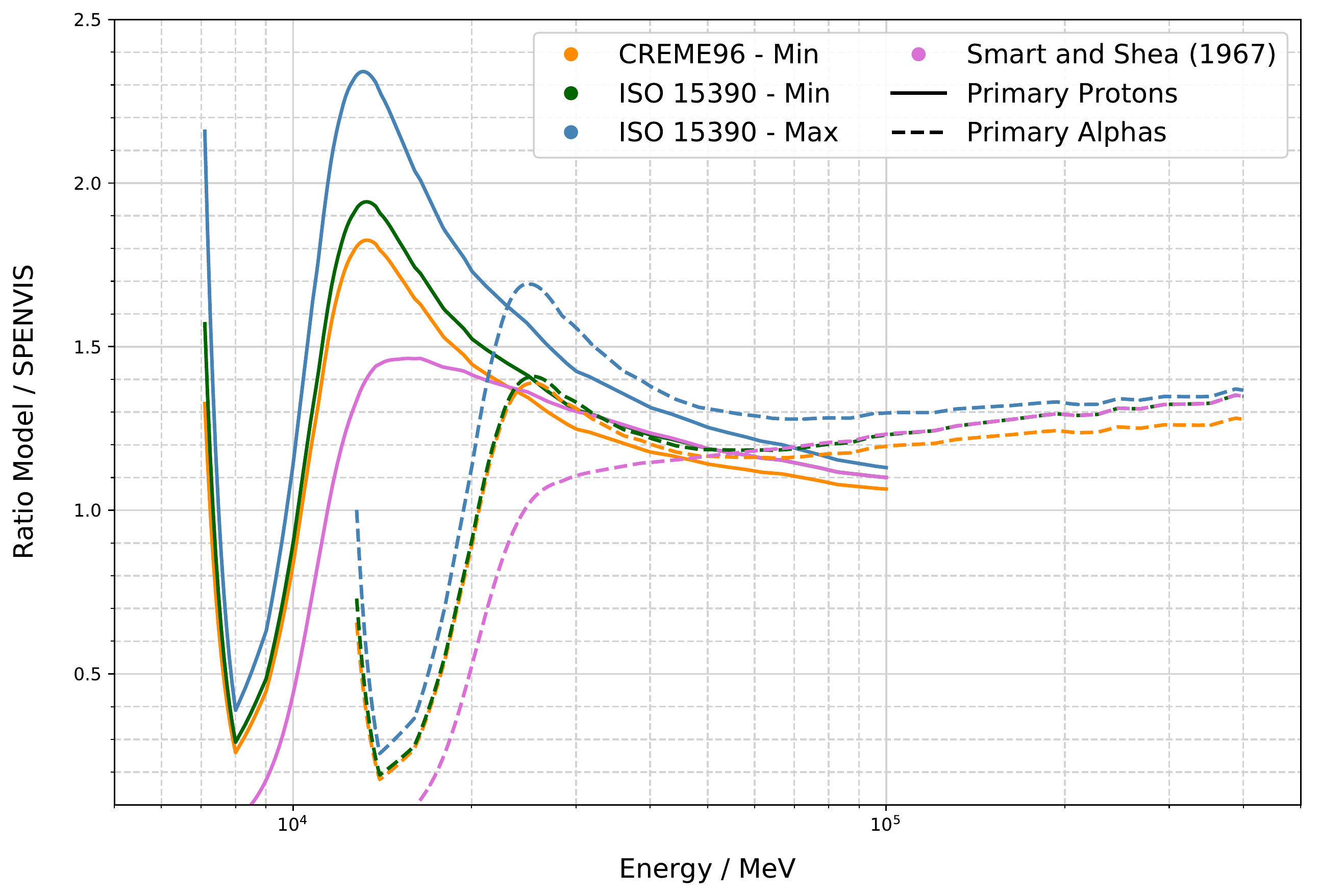}
\caption{Ratio between the model presented here and the results by SPENVIS, obtained using different models for the galactic cosmic rays (CREME96 and ISO 15390) in different solar activity condition (minimum and maximum) and two different methods for computing the rigidity cut-off (St\o rmer with eccentric dipole model, the default option, and Smart and Shea (1967) with the IGRF model, the latter of which was used only with the CREME96 model).}
\label{fig:spenvis}
\end{center}
\end{figure}

Cosmic rays consist mainly of protons, a lot of which are trapped by the geomagnetic field and never reach LEO. AMS data (\cite{PhysRevLett.114.171103}) for the unmodulated proton spectrum have been used, together with the reduction factor from the geomagnetic cutoff described in \cite{0004-637X-614-2-1113}, to model the component:
\begin{equation}\label{eq:primprot}
    F = F_{p, AMS} \cdot \frac{1}{1+(R/R_{cutoff})^{-12}}
\end{equation}
where $R$ is the proton rigidity and $R_{cutoff}$ is defined in equation \ref{eq:Rcut}. Results are shown in Fig. \ref{fig:prcharged}.\\
While results obtained by tools like ESA's SPace ENVironment Information System (SPENVIS\footnote{\url{https://www.spenvis.oma.be/}}) are more realistic, by considering, e.g., the variation of $R_{cutoff}$ along the orbit, the approach described by eq. \ref{eq:primprot} was chosen for consistency with the approach used to calculate the primary electrons/positrons spectra (that can not be calculated using SPENVIS), and to allow the use of a single tool to calculate all the background components for different LEOs. A comparison with the results from SPENVIS for an equatorial orbit at 550 km is shown in Fig. \ref{fig:spenvis}. The results are shown only in the energy range provided by SPENVIS (1~MeV$ /n <E<10$0~GeV$ /n$) and are calculated using the two available cosmic-ray models: CREME96 for a solar minimum and ISO 15390 for both minimum and maximum solar activity. When using the default description of the geomagnetic field, the differential spectrum is null for 1 MeV$ < E \lesssim 7$GeV in case of protons, and 1 MeV$/n < E \lesssim 3$GeV$ /n$ for the alpha particles. Differences up to a factor of 2 are visible around the value of the geomagnetic cutoff. By changing the method used by SPENVIS to compute the rigidity cut-off\footnote{\url{https://www.spenvis.oma.be/help/models/magshielding.html}} the ratio with the presented model reaches a maximum factor of 1.5 at the energy of the cutoff. For energies higher than the cutoff SPENVIS predicts a higher flux. It should be noted that such differences have little effect on the detector activation due to the creation of unstable isotopes, which is dominated by the higher part of the spectrum.

\subsection{Primary Alpha Particles}
A similar approach as the one used for the protons has been used also for the primary alpha particles. The unmodulated spectrum was derived from AMS data (\cite{PhysRevLett.115.211101}), to which the reduction factor from the geomagnetic cutoff was lately applied:
\begin{equation}
    F = F_{\alpha, AMS}  \cdot \frac{1}{1+(R/R_{cutoff})^{-12}}
\end{equation}
where $R$ is the particle rigidity and $R_{cutoff}$ is defined in equation \ref{eq:Rcut}.\\
Again, as per the primary protons, while the SPENVIS results are more realistic, the described approach was followed for consistency. As per the protons, and with similar results, a comparison with the results from SPENVIS for an equatorial orbit at 550 km is shown in Fig. \ref{fig:spenvis}.

\subsection{Primary Electrons}

Consistently with what has been done for the protons and alpha particles, the primary spectrum from AMS data (\cite{PhysRevLett.113.121102}) is used, with the subsequent application of the geomagnetic cutoff modulations:
\begin{equation}\label{eq:primele}
    F = F_{e, AMS}  \cdot \frac{1}{1+(R/R_{cutoff})^{-6}}
\end{equation}
Although the geomagnetic cutoff is expected to be the same regardless of the particle type, a reduction factor of the form $1/1+(R/R_{cutoff})^{-n}$, with $n=6$, gives a better description of the AMS-01 (\cite{200010}) data for electrons and positrons than with $n=12$ (see \cite{0004-637X-614-2-1113}). The resulting flux is shown in Fig. \ref{fig:prcharged}.

\subsection{Primary Positrons}

Exactly like for the primary electrons, a combination of AMS data (\cite{PhysRevLett.113.121102}) and the subsequent application of the same geomagnetic reduction factor with $n=6$ (eq. \ref{eq:primele}), were used. The results are shown in Fig. \ref{fig:prcharged}.

\subsection{Secondary Protons}

\begin{figure}[ht]
\begin{center}
\includegraphics[width=0.95\textwidth]{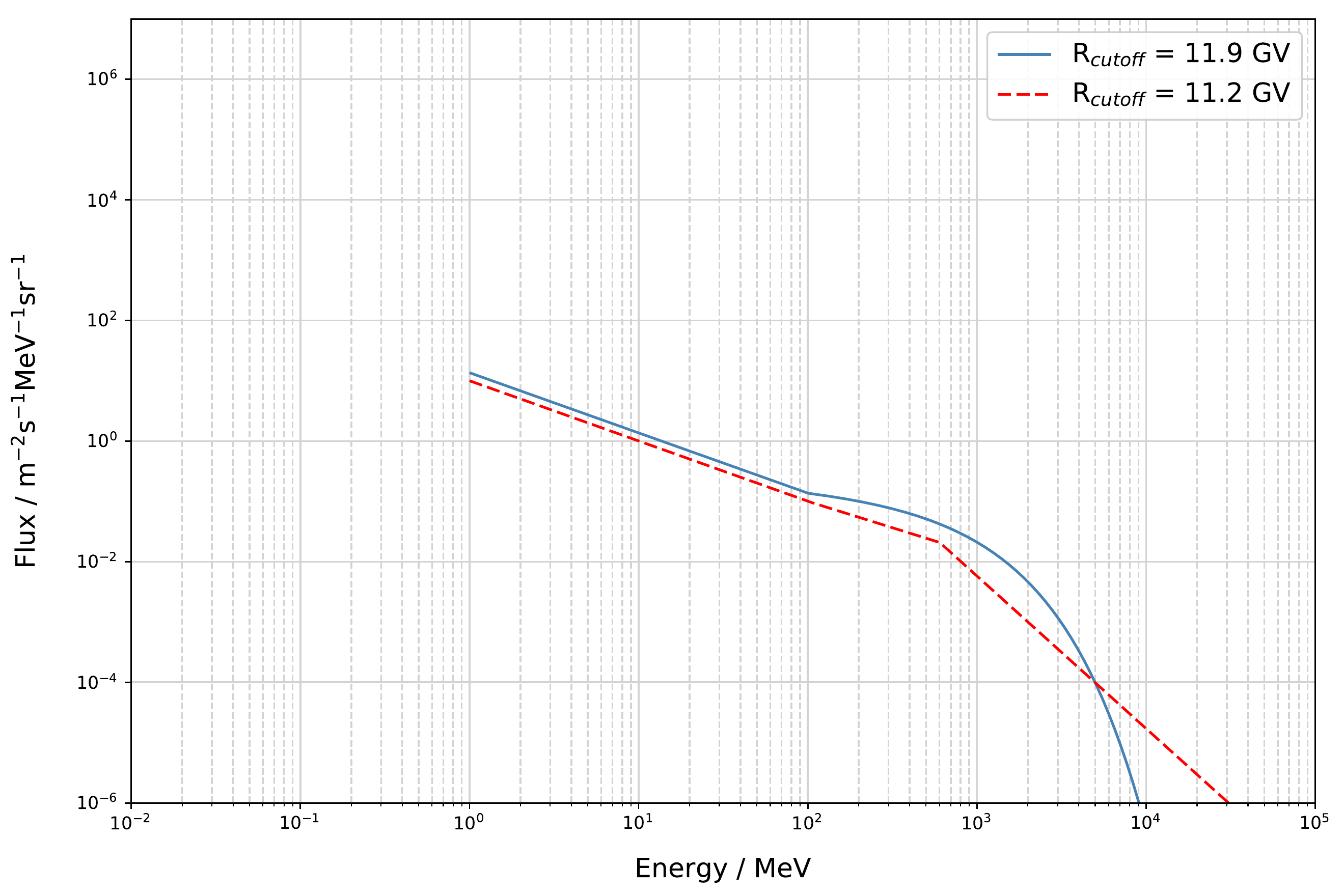}
\caption{Spectra of on-orbit secondary protons describing both the upward and downward component for two different geomagnetic cutoffs: $R_{cutoff}=11.9$ GV (the one, e.g. for a 550 km equatorial orbit) and $R_{cutoff}=11.2$ GV (e.g., for a 550 km orbit at an inclination of 10$^\circ$).}
\label{fig:secprotons}
\end{center}
\end{figure}

When cosmic rays interact with the atmosphere's particles, they produce secondary particles. The secondary protons, shown in Fig. \ref{fig:secprotons}, are computed following \cite{0004-637X-614-2-1113}, based on AMS-01 data (\cite{2000215}):
\begin{itemize}
\item 1 MeV $< E <$ 100 MeV: Equation 9 of \cite{0004-637X-614-2-1113} \begin{equation}\label{eq:secpr1}
    F= \frac{0.136}{10^7}\cdot\left(\frac{E}{100 ~\mathrm{MeV}} \right)^{-1}
\end{equation}
is extrapolated down to 1 MeV. Below 1 MeV there is no probability for a proton of creating an unstable isotope which decay could add to the total background. 
\item E $>$ 100 MeV: \begin{equation}\label{eq:secpr2}
    F= \frac{0.123}{10^7}\cdot\left(\frac{E}{1000 ~\mathrm{MeV}} \right)^{-0.155}\exp{-\left(\frac{E}{510 ~\mathrm{MeV}} \right)^{-0.845}}
\end{equation}
\end{itemize}
\cite{0004-637X-614-2-1113} model the secondary proton flux at the altitude of AMS-01 (380 km) as a function of the geomagnetic latitude $\lambda$. We assume that the geomagnetic cutoff $R_{cutoff}$ is a more relevant parameter to describe the particle flux for different altitudes and inclinations. Equations \ref{eq:secpr1} and \ref{eq:secpr2} describe both the upward and downward fluxes of the secondary protons in high geomagnetic cutoff regions ($R_{cutoff}\gtrsim 11.5$ GV). At lower cutoff values, down to $\sim 10.4$ GV, the spectra of the downward and upward components take the shape of a broken power-law at energies greater than 100 MeV (\cite{0004-637X-614-2-1113}):
\begin{itemize}
\item 100 MeV $< E <$ 600 MeV: 
\begin{equation}
    F= \frac{0.1}{10^7}\left(\frac{E}{100 ~\mathrm{MeV}}\right)^{-0.87}
\end{equation}
\item $E \geqslant$ 600 MeV:
\begin{equation}
    F= \frac{0.1}{10^7}\left(\frac{600}{100}\right)^{-0.87} \left(\frac{E}{600 ~\mathrm{MeV}}\right)^{-2.53}
\end{equation}
\end{itemize}
The two components assume different shapes at lower cutoff values (see \cite{0004-637X-614-2-1113}).

\subsection{Secondary Electrons and Positrons}
\begin{figure}[ht]
\begin{center}
\includegraphics[width=0.95\textwidth]{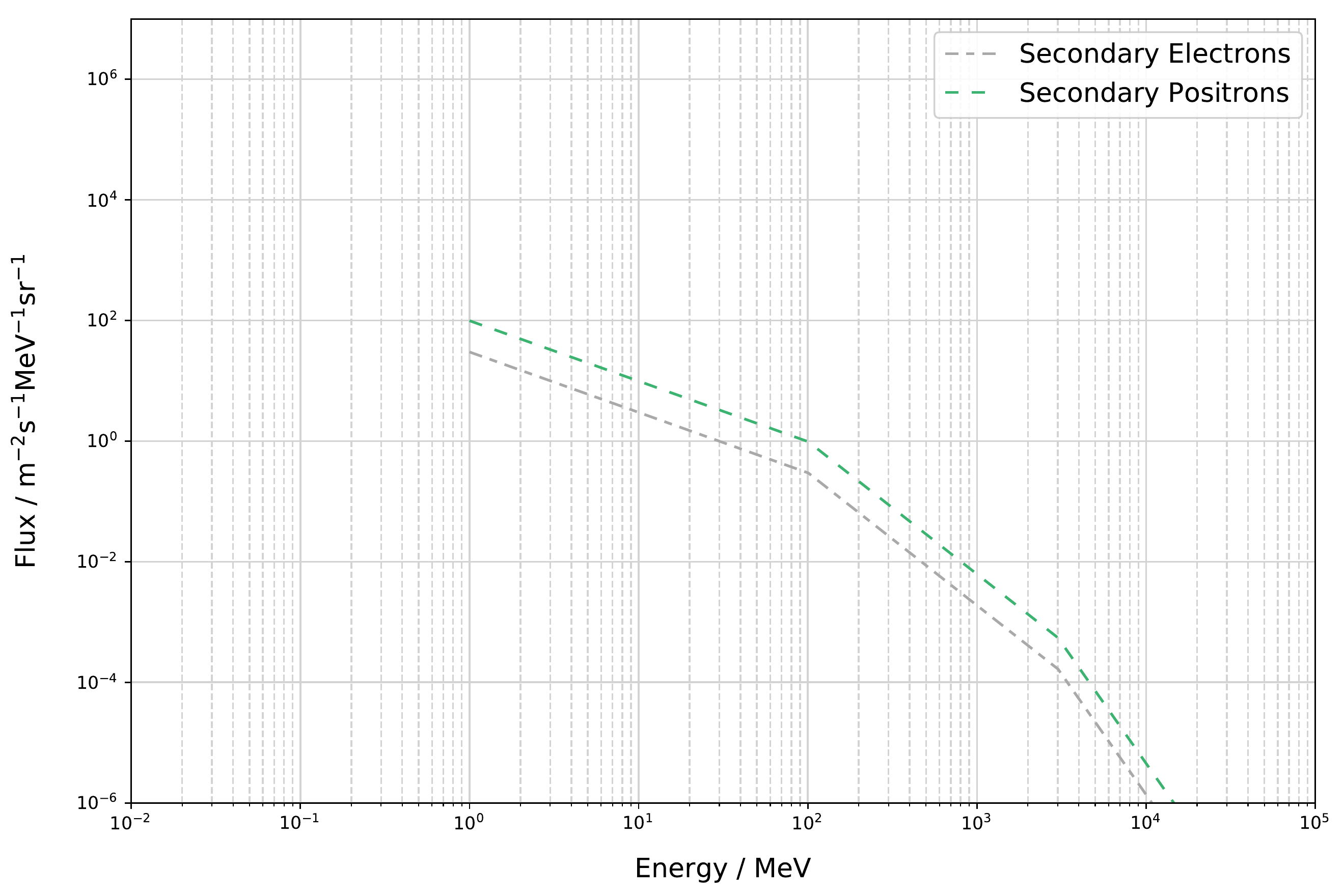}
\caption{Secondary electrons and positrons spectra.}
\label{fig:elecpos}
\end{center}
\end{figure}
The spectra of the secondary electrons and positrons, shown in Fig. \ref{fig:elecpos}, are based on data from AMS-01 (\cite{200010}), modeled as in \cite{0004-637X-614-2-1113}, extrapolating their equations to 1 MeV. For the electrons the equations are:
    \begin{itemize}
    \item 1 MeV $< E <$ 100 MeV: \begin{equation}
    F= \frac{0.3}{10^7}\cdot\left(\frac{E}{100 ~\mathrm{MeV}} \right)^{-1}
    \end{equation}
\item 100 MeV $< E <$ 3 GeV: \begin{equation}
    F= \frac{0.3}{10^7}\cdot\left(\frac{E}{100 ~\mathrm{MeV}} \right)^{-2.2}
    \end{equation}
\item $E>$ 3 GeV\begin{equation}
    F= \frac{0.3}{10^7}\cdot\left(\frac{E}{100 ~\mathrm{MeV}} \right)^{-2.2}\left(\frac{E}{3 ~\mathrm{GeV}} \right)^{-4}
\end{equation}
\end{itemize}
For the positrons the same equations are multiplied by a factor 3.33 from the e$^+/$e$^-$ ratio in equatorial orbit. As per the secondary protons, the spectra of the secondary electrons and positrons depends on the geomagnetic cutoff. In particular, the equations above are valid down to a cutoff value of $\sim 10.4$ GV, corresponding to an inclination of $\sim 15^\circ$ at an altitude of 550 km.

\subsection{Atmospheric Neutrons}
\begin{figure}[ht]
\begin{center}
\includegraphics[width=0.95\textwidth]{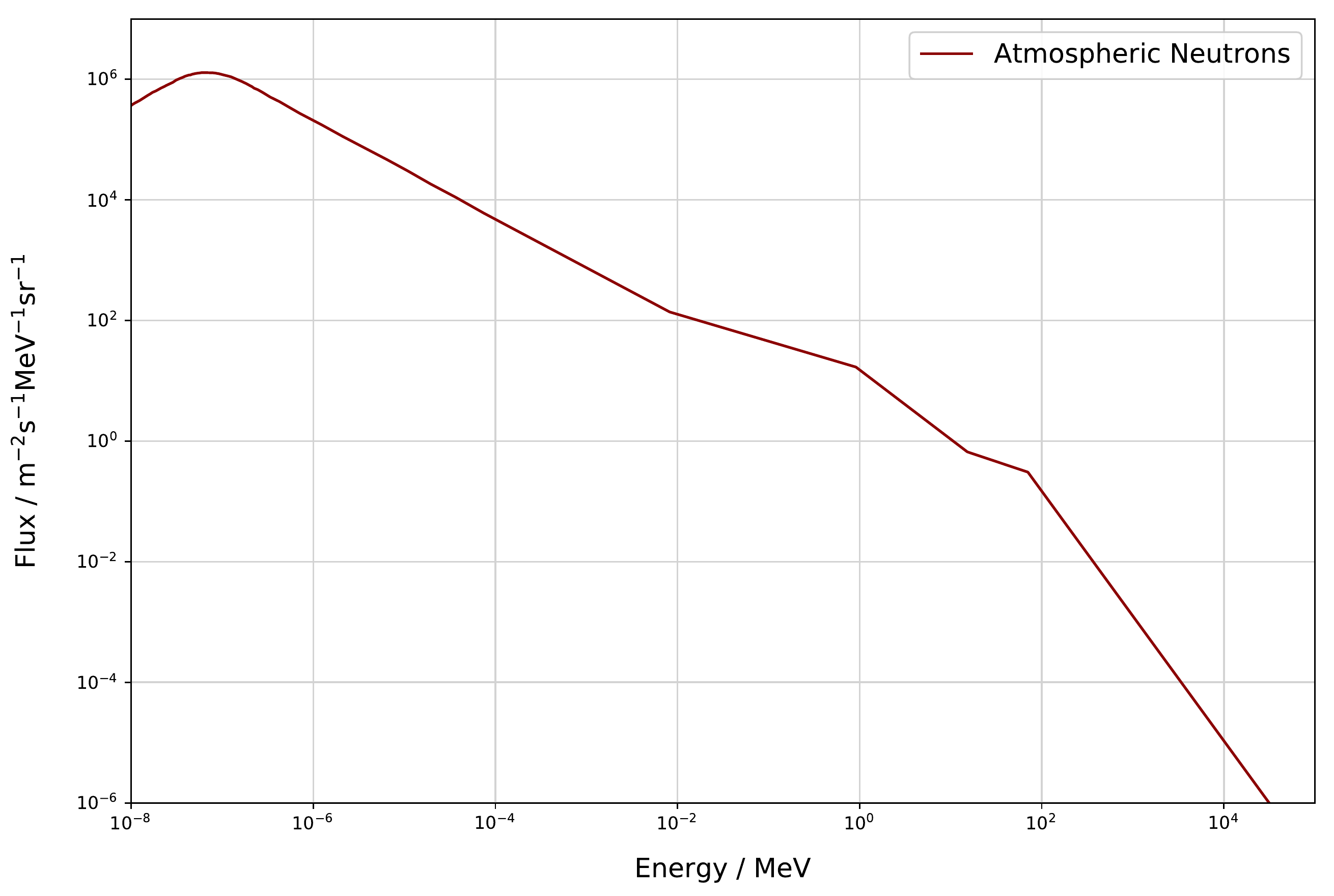}
\caption{On-orbit spectrum (out of the atmosphere) of the cosmic-ray induced atmospheric neutrons.}
\label{fig:neutrons}
\end{center}
\end{figure}
When interacting with the atmosphere, galactic cosmic-rays can create hadronic showers. Neutrons created in such showers can reach instrument on LEOs.\\ 
As reported in \cite{2015APh....62..230K}, since the downward component tends to zero at high altitudes, the on-orbit atmospheric neutrons spectrum is calculated following the description of only the upward component.\\
The flux can be described by a power-law $F=x_i \cdot E ^{-y_i} \cdot 10^{-3}$ with four different normalization and exponents in four different energy intervals (\cite{2015APh....62..230K}). The equations describing $x_i$ and ${y_i}$ depend on the geomagnetic latitude $\lambda$ and the solar activity parameter $S=(\phi  - 250~\mathrm{MV})/(859~\mathrm{MV})$, where $\phi$ is the solar modulation potential ($S=0.47$ for $\phi=650$ MV). These equations are:

\begin{itemize}
    \item E $<$ 0.9 MeV: \begin{equation}
\begin{split}
        &a = 3 \cdot 10^{-4} + (7 -5\cdot S)\cdot 10^{-3} [1-\tanh(180-4\cdot \lambda)] \\
        &b = 1.4 \cdot 10^{-2} + (1.4 -0.9\cdot S)\cdot 10^{-1} [1-\tanh(180-3.5\cdot \lambda)] \\
        &c = 180  - 42 \cdot [1-\tanh(180-5.5\cdot \lambda)] \\
        &d = -8 \cdot 10^{-3} + (6 - S)\cdot 10^{-3} [1-\tanh(180-4.4\cdot \lambda)] \\
        &x_1= (a\cdot p + b) e^{-p/c} + d\\
        &y_1 = -0.29 e^{-p/7.5}+0.735
\end{split}
\end{equation}
    \item 0.9 MeV $<$ E $<$ 15 MeV:  \begin{equation}
\begin{split}
        &x_2= x_1 \cdot 0.9^{-y_1+y_2}\\
        &y_2 = -0.247 e^{-p/36.5}+1.4
\end{split}
\end{equation}
    \item 15 MeV $<$ E $<$ 70 MeV: \begin{equation}
\begin{split}
        &x_3= x_2 \cdot 15^{-y_2+y_3}\\
        &y_3 = -0.4 e^{-p/40}+0.9
\end{split}
\end{equation}
    \item E $>$ 70 MeV:\begin{equation}
\begin{split}
        &x_4= x_3 \cdot 70^{-y_3+y_4}\\
        &y_4 = -0.46 e^{-p/100}+2.53
\end{split}
\end{equation}
\end{itemize}
The results are shown in Fig. \ref{fig:neutrons}. As shown in \cite{2015APh....62..230K}, this model is found to be in good agreement with data from high altitude aircraft, balloon-borne experiments and previously published simulations.\\
The model developed by \cite{2015APh....62..230K} is valid down to 8~keV. The spectrum at the top of the atmosphere from \cite{JZ068i020p05633}, scaled to the \cite{2015APh....62..230K} results, was used to describe the spectrum between 0.01~eV and 8~keV. The \cite{JZ068i020p05633} results outside the atmosphere for energies lower than 1 MeV are compatible with the calculations in \cite{JA078i016p02715}, which in turns are compatible with the results in \cite{2015APh....62..230K}.\\
To compute the result shown in Fig. \ref{fig:neutrons}, the result of the previously described equations has been divided by $\Omega = 4\pi - 2\pi (1-cos \theta _H) = 2\pi (1+cos \theta _H)$ to get a flux per unit steradian. \\

\subsection{Total Spectrum}\label{s:totalsp}
\begin{figure}[p]
\begin{center}
\includegraphics[width=0.95\textwidth]{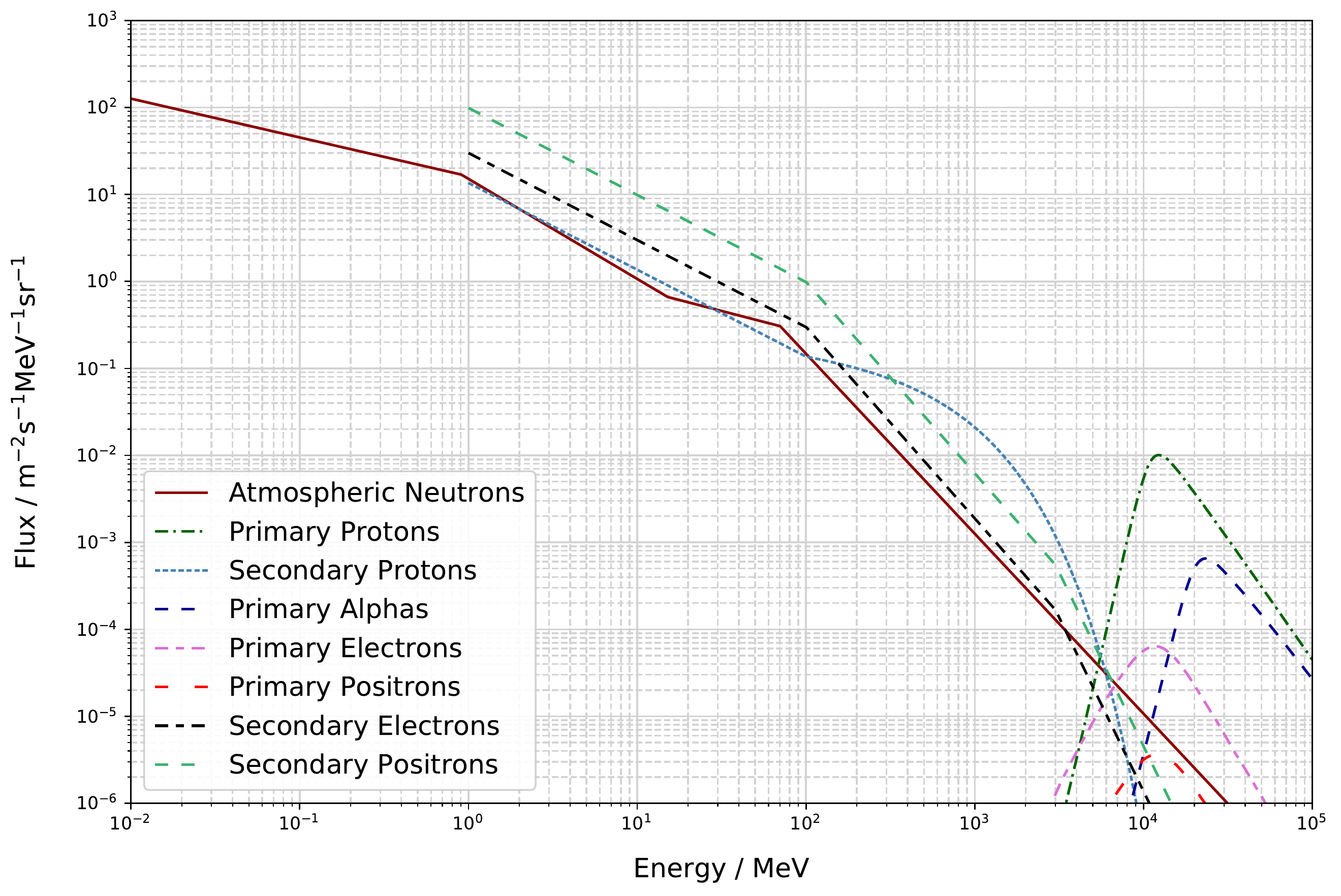}
\includegraphics[width=0.95\textwidth]{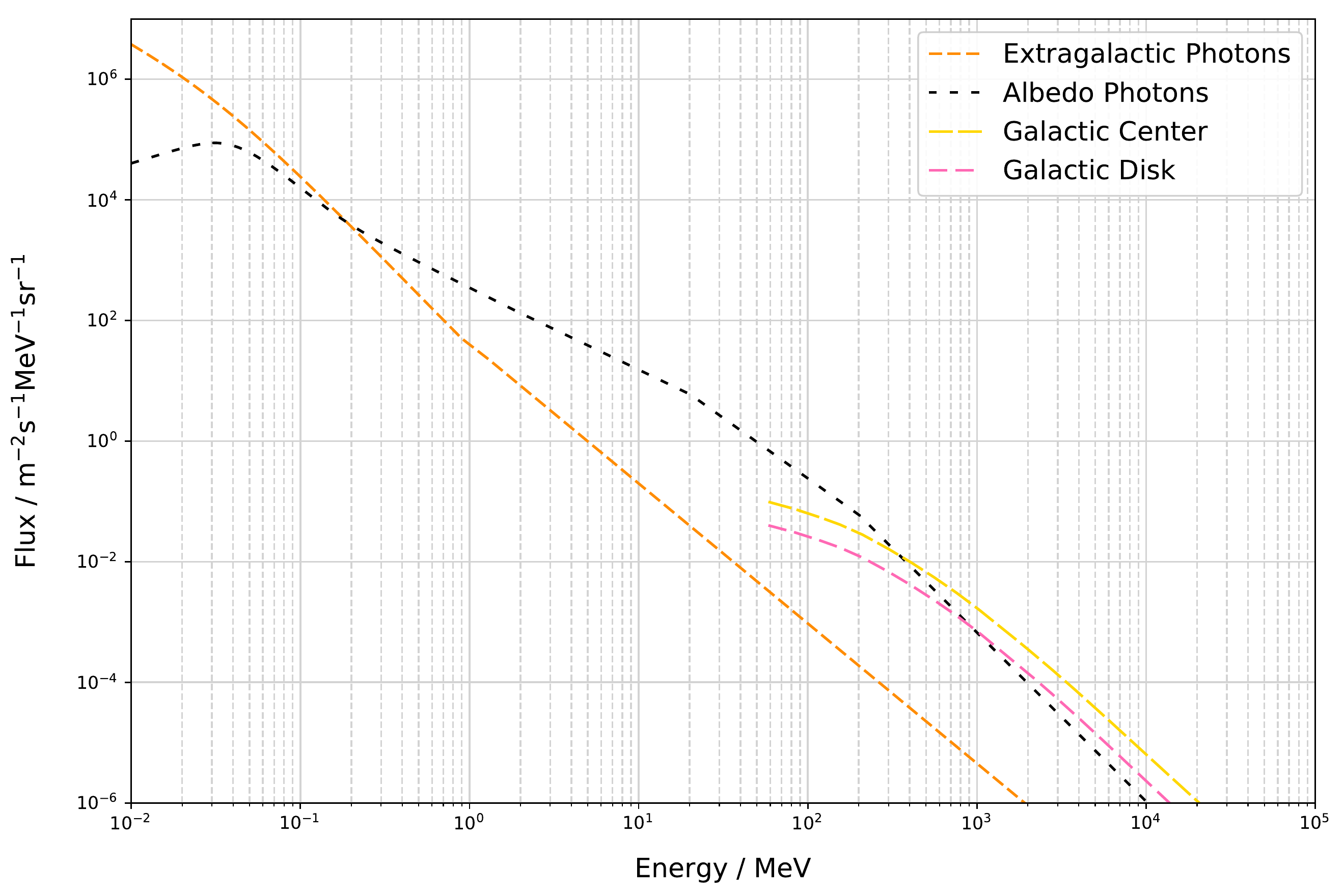}
\caption{Full on-orbit background spectrum for charged particles and neutrons (\textit{top}) and photons (\textit{bottom}). All spectra components shown are divided by the solid angle of their region of origin. The pointing and field-of-view of an hypothetical instrument are therefore not considered}
\label{fig:fullspectrum}
\end{center}
\end{figure}

\begin{figure}[ht]
\begin{center}
\includegraphics[width=0.95\textwidth]{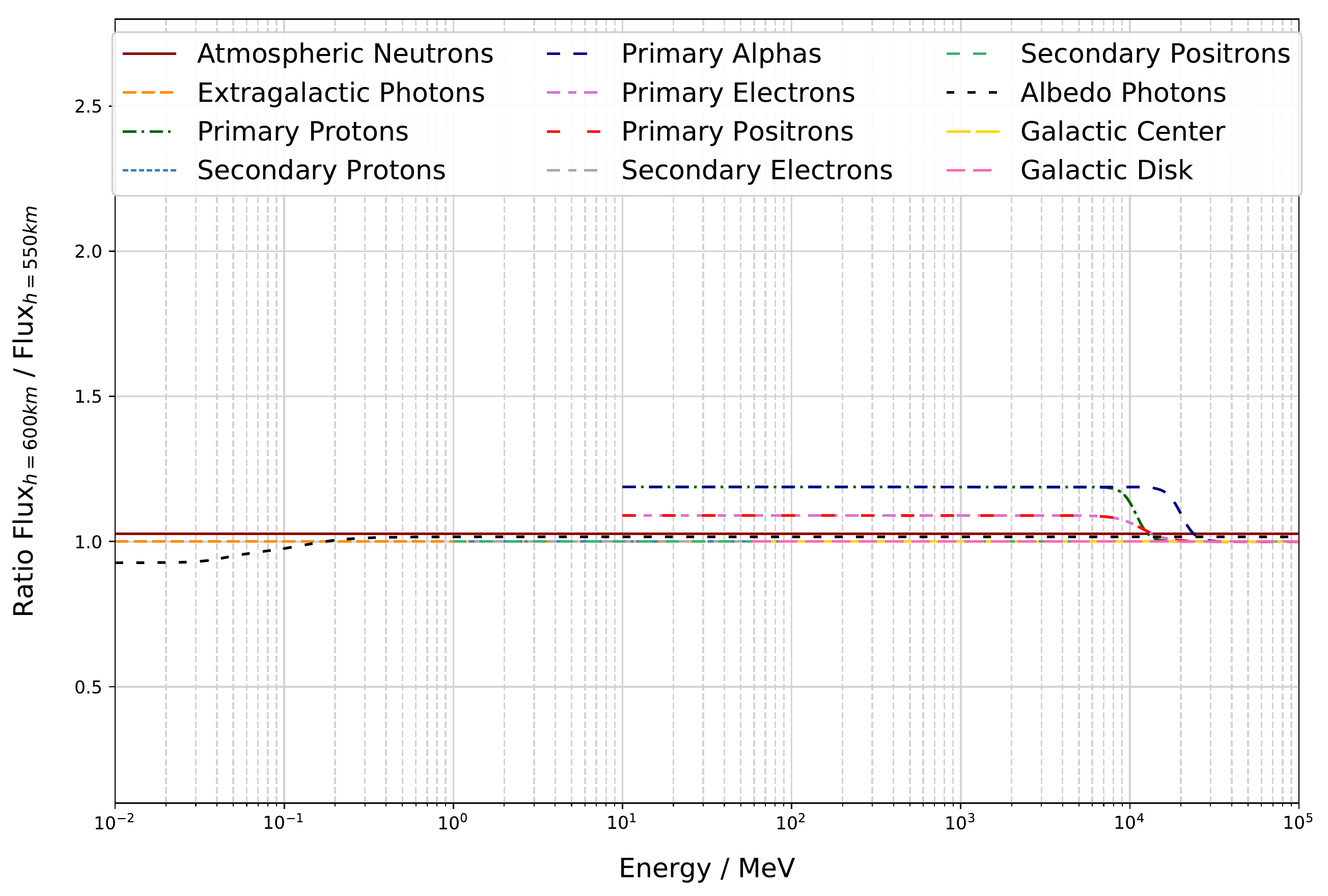}
\includegraphics[width=0.95\textwidth]{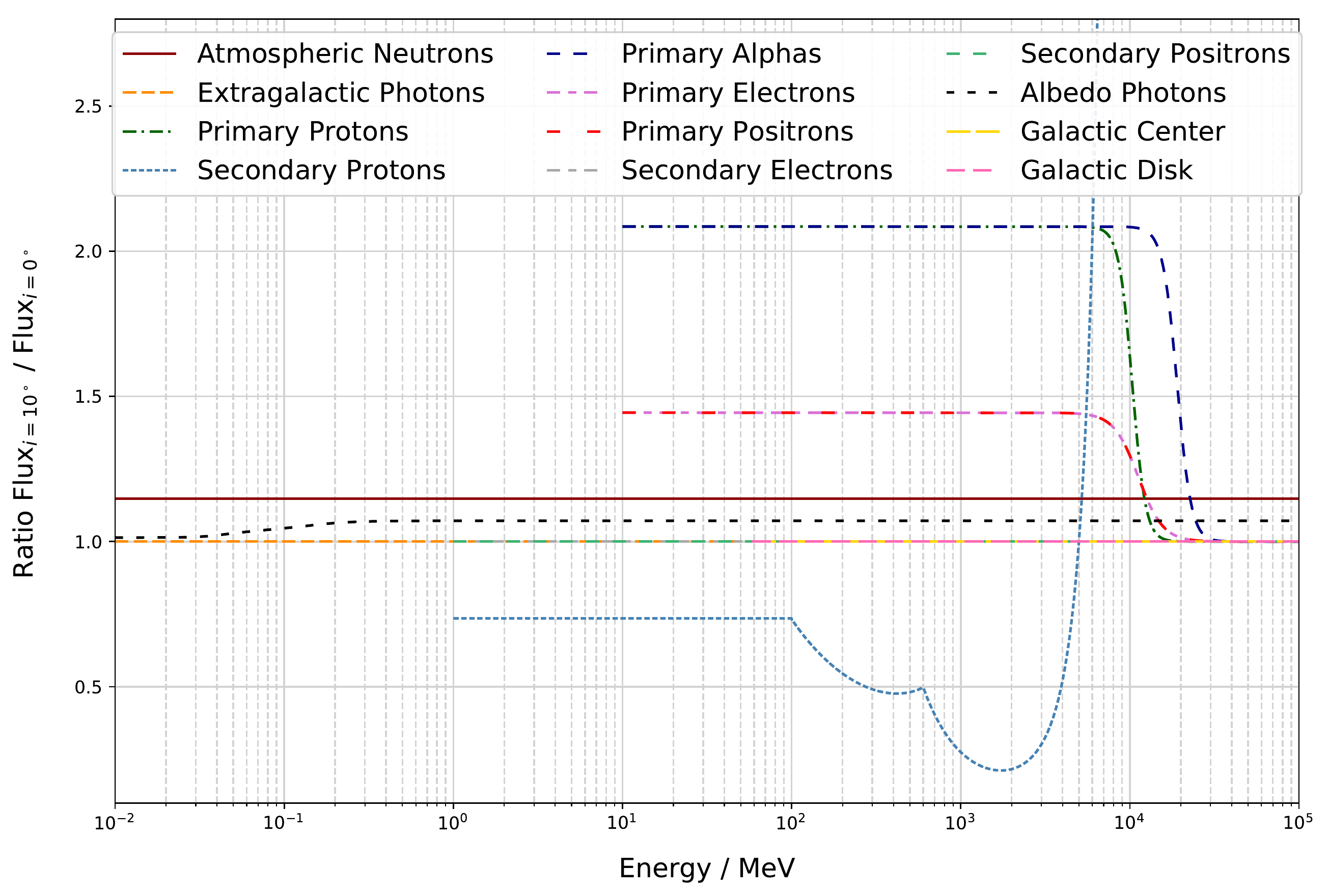}
\caption{Ratio between fluxes obtained for different orbits: at the variation of the altitude, from 550 km to 600 km (\textit{top}) or inclination, from $0^\circ$ to $10^\circ$ (\textit{bottom}).}
\label{fig:incaltchange}
\end{center}
\end{figure}

A collection of all the results for a 550 km equatorial orbit is shown in Fig. \ref{fig:fullspectrum}. Up to $\sim100$ keV, the dominant component of the photon background spectrum is represented by photons of extragalactic origins, subsequently overtaken by gamma rays generated in the atmosphere. Depending on the line of sight, the main component of the background can become the Galactic center, starting from $\sim400$ keV, or the Galactic disk, starting from 1~GeV, with the albedo photons being still a non negligible component up to the highest energy. \\
Given the big uncertainty due to a lack of data and the variation due to the solar modulation, for energies lower than 1~MeV only the neutron component is calculated. Secondary charged particles, positrons first and protons later, give the highest contribution up to $\sim6$~GeV, where they are surpassed by the primary protons. While most of the particles interaction with the satellite should be rejected by an anticoincidence system (99.97\% in the case of Fermi \cite{2009ApJ...697.1071A}), some of them could interact with part of the detector and create long living isotopes which decay could resemble a Compton event. A study on the background activation, using the spectrum of Fig. \ref{fig:fullspectrum}, is presented in the next section.\\
At the change of either altitude or orbit inclination, both shown in Fig. \ref{fig:incaltchange}, the spectra of some of the components remain unchanged. The photons background, either Galactic or extra-Galactic, does not change in any way with the orbit, regardless of the model used. The spectra of secondary charged particles (protons, electrons, and positrons), as modeled here, depends solely on the geomagnetic cutoff which, as demonstrated before, varies more with the inclination than with the altitude (see Fig.\ref{s:model}). For a small change of altitude no changes are seen in any of the secondary components. The secondary protons spectra however change its shape at higher inclination leading to a decrease in the spectrum up to a factor $>2$ from $\sim$300~MeV up to $\sim 4$~GeV, with a subsequent rapid increase. For such a change in inclination (0$^\circ$ to 10$^\circ$), the secondary electrons and positrons spectra remain unchanged.
The albedo photons spectrum as shown in Fig. \ref{fig:photons} and \ref{fig:fullspectrum} changes both with the inclination and the altitude of the orbit. The variation is due to the dependence on the geomagnetic cutoff and, at low energies, a more complicated dependence on the Earth viewing angle. A difference of $10\%$ or less is found when changing to 600 km maintaining the same inclination, or to $10^\circ$ keeping the altitude constant. Up to 200~keV, in the energy range dominated by the reflected extragalactic X-ray background, dependent only on the Earth viewing angle, a different trend with up to 7\% reduction at increasing altitude can be observed.\\
On LEOs, the atmospheric neutrons spectrum is more dependent on the inclination than on the altitude. It changes by $\sim17\%$ at the variation of the inclination, and less than 10\% when changing altitude.\\
All primary charged particles spectra are dependent on both altitude and inclination of the orbit through their dependence on the geomagnetic cutoff $R_{cutoff}$. The differences in the spectra thus appear at energies lower than the cutoff, quickly disappearing at the highest energy. For protons and alpha particles, the spectrum of which is modulated by a factor $R_{cutoff} ^{-12}$, this difference can reach a factor of 2 increasing the inclination or 20\% while changing the altitude. Electrons and positrons spectra are more loosely dependent on the geomagnetic cutoff, only by a factor $R_{cutoff}^{-6}$, leading to a maximum difference of a factor $\sim1.4$. However these differences appear in an energy range where the particle fluxes are strongly reduced by the geomagnetic modulation, making them not important in the background calculation.
\begin{table}[t]
    \centering
    \begin{tabular}{l c r}
        & Orbit Parameters & Energy\\\hline
        Extragalactic Photons & Independent & 4 keV - 820 GeV\\
        Galactic Photons & Independent & 58 MeV -  $\sim 513$ GeV\\
        Albedo Photons & All LEOs & 1 keV - 400 GeV\\
        Primary Protons & All LEOs & 10 MeV - 10 TeV\\
        Primary Alphas & All LEOs & 10 MeV - 10 TeV\\
        Primary Electrons/Positrons & All LEOs & 570 MeV - 429 GeV\\
        Secondary Protons & $1.06\leq R_{cutoff}\leq 12.47$ & 1 MeV - 10 GeV\\
        Secondary Electrons/Positrons & $1.06\leq R_{cutoff}\leq 12.47$ & 1 MeV - 20 GeV\\
        \multirow{2}{*}{Atmospheric Neutrons} & Altitude: $\sim100$ km - $\sim1000$ km & \multirow{2}{*}{0.01 eV - $\sim 30$ GeV}\\
         & Inclination: $<65^\circ$ & \\
         
    \end{tabular}
    \caption{Validity limits of the different components of the background.}
    \label{tab:limits}
\end{table}
The validity limits of the model for the different components of the background are presented in Tab. \ref{tab:limits}.

\section{South Atlantic Anomaly}\label{s:SAA}

\begin{figure}[ht]
\begin{center}
\includegraphics[width=0.95\textwidth]{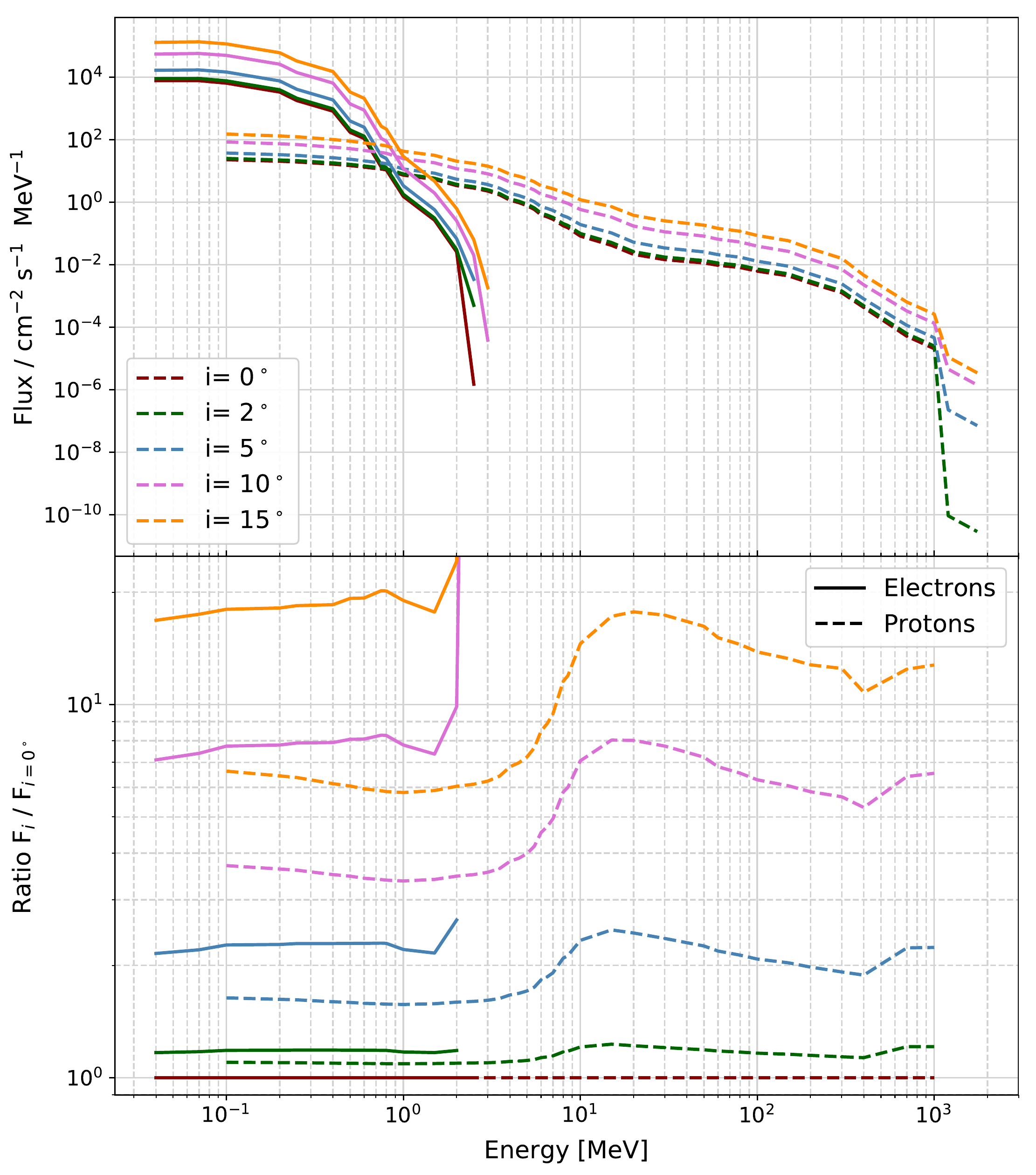}
\caption{\textit{Top}: Average particle flux experienced in one month by a satellite passing through the South Atlantic Anomaly, as calculated by the AE9/AP9-IRENE model. \textit{Bottom}: Ratio of the spectrum as calculated for different inclinations with respect to the spectrum on an equatorial orbit ($i=0^\circ$)}
\label{fig:SAAflux}
\end{center}
\end{figure}
\begin{figure}[ht]
\begin{center}
\includegraphics[width=0.95\textwidth]{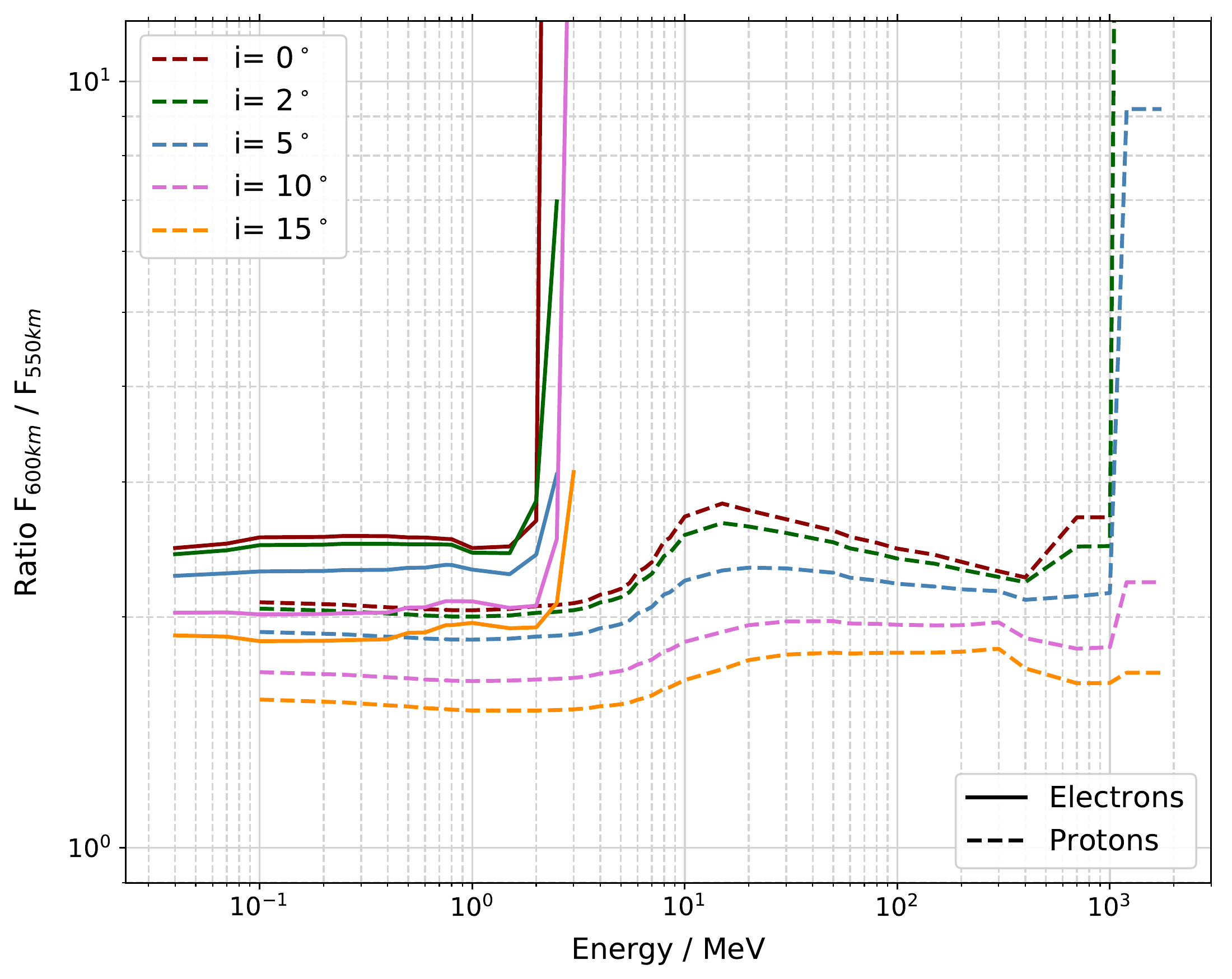}
\caption{Ratio of the trapped particles flux at different altitudes calculated for different orbit inclinations. Similarly to Fig. \ref{fig:SAAflux}, the different extensions of the spectra at higher energies create a sharp increase in the ratio. It should be noted that at these energies the fluxes are already negligible.}
\label{fig:SAAAltitude}
\end{center}
\end{figure}

The South Atlantic Anomaly (SAA) is the region where the inner Van Allen radiation belt comes closest to the Earth surface. Satellites passing through this area are subject to intense radiation, caused mainly by protons and electrons. Scientific observations are usually stopped while passing through the SAA, although some diagnostic data can still be recorded (see e.g. \cite{2009APh....32..193A}). It is possible for protons, interacting with the satellite, to create short and long lived radioisotopes, which would decay after resuming observation, adding to the total background. To simulate the creation and decay of such isotopes, the first step is to estimate the SAA spectrum along a given orbit.\\
The AE9/AP9-IRENE (\cite{Ginet2013}) models (version 1.5, \cite{8101531}) have been used to simulate the SAA passage at an altitude of 550 km and along differently inclined ($0^\circ,1^\circ,2^\circ,3^\circ,4^\circ,5^\circ,10^\circ$) orbits. The model allowed to compute the mean, omnidirectional, flux along the orbit with a 10 s time step. The results for a total simulated time of 1 month ($\sim455$ orbits at 550 km) are shown in Fig. \ref{fig:SAAflux}, together with a ratio of the values at different inclinations. The average flux increases with the orbit inclination, in a similar way for both electrons and protons. The difference with respect to the equatorial orbit reaches a factor of more than 6 for an inclination of $10^\circ$, and about an order of magnitude for $i=15^\circ$. At greater inclinations, a greater difference between the low ($<10$ MeV) and high energy part of the proton spectrum can also be noted.\\
With an increase of altitude the flux increases, more at lower orbit inclinations than at higher ones, as shown in Fig. \ref{fig:SAAAltitude}. To an increase in altitude of 50 km, it corresponds an increment of the flux between a factor of $\sim2$ and $\sim3$ for inclinations lower than $5^\circ$.\\
A low inclination and low altitude orbit ensures the lowest possible flux during the passage through the SAA. A perfectly equatorial orbit would grant the best conditions, but little changes are observed for inclination smaller than 3$^\circ$. To this kind of orbit corresponds also the smallest average time passed by the satellite inside the SAA. For a 550 km equatorial orbit, considering particles with $E>0.1$ MeV, the time corresponds to around 18\% of the orbit period (95 minutes at 550 km). An increase larger than 20\% is seen for a 600 km orbit, while differences up to 20\% are observed with varying inclinations.

\section{Activation Simulations}\label{s:AcSim}
\begin{table}[t]
    \centering
    \begin{tabular}{l c c c c r}
        \multirow{3}{*}{Isotope} & \multirow{3}{*}{T$_{1/2}$} & \multicolumn{3}{c}{Activation$_i$ / Activation$_{max}$} & \multirow{3}{*}{Volumes}\\\cline{3-5}
         &  & General & \multicolumn{2}{c}{SAA} &  \\
         &  & Background & 1 yr & 1 orb. &  \\
        \hline
        \multirow{4}{*}{$^{11}$C} & \multirow{4}{*}{20 m} &  \multirow{4}{*}{0.2 (0.6 sp)} & \multirow{4}{*}{1} & \multirow{4}{*}{0.8} & Anticoincidence\\
        &  &  & & & Supports \\
        &  &  & & & Spacecraft \\
       &  &  & & & Electronics\\\hline
        &  &  & & & \multirow{11}{*}{Calorimeter Crystals} \\
         $^{128}$I & 25 m &  1 & 0.9 & 0.4 & \\
         $^{126}$I & 13 d & 0.3 & 0.7 & $<0.1$ &  \\
         $^{132}$Cs & 6.5 d & 0.2 & 0.8 & $<0.1$ &  \\ 
         $^{134}$Cs & 2.1 yr & 0.5 & 0.3 & $<0.1$ & \\ 
         $^{125}$I & 59 d & 0.2 & 0.5 & $<0.1$ &  \\ 
         $^{131}$Cs & 9.7 d & 0.1 & 0.5 & $<0.1$ & \\
         $^{123}$I & 13 h  & 0.1 & 0.5 & $<0.1$ & \\ 
         $^{122}$I & 3.6 m & 0.1 & 0.1 & 0.1 &  \\
         $^{126}$Cs & 1.6 m & $<0.1$ &  $<0.1$ &  0.1  &  \\
         & &  & \\\hline
        
         \multirow{4}{*}{$^{25}$Al} & \multirow{4}{*}{7.2 s} & \multirow{4}{*}{$<0.1$ (0.1 sp)} & \multirow{4}{*}{$<0.1$} & \multirow{4}{*}{0.5} & Spacecraft\\
         &  &  & & & Electronics\\
       &  &  & & & Tracker Wafers\\
      &  &  & & & Calorimeter Diodes\\\hline
        
        \multirow{3}{*}{$^{28}$Al} & \multirow{3}{*}{2.2 m} & \multirow{3}{*}{0.3 (0.4 n)} & \multirow{3}{*}{$<0.1$} & \multirow{3}{*}{0.5} & Calorimeter Diodes\\
       &  &  & & &  Electronics\\
        &  &  & & & Tracker Wafers\\\hline
        
         \multirow{3}{*}{$^{15}$O} & \multirow{3}{*}{2 m} & \multirow{3}{*}{0.1 (0.3 sp)} & \multirow{3}{*}{$<0.1$} & \multirow{3}{*}{1} & Anticoincidence\\
       &  &  & & & Electronics\\
       &  &  & & & Supports\\
    \end{tabular}
    \caption{List of the most relevant isotopes created by the particles passing through the different sub-systems of the e-ASTROGAM satellite model. Results are given as relative to the most active source of background, i.e. $^{128}$I for the general background, $^{11}$C for SAA-1 yr and $^{15}$O for the SAA short-term activation. The results are generally consistent for all the components, some exception to this are observed for the atmospheric neutrons (n) and secondary protons (sp).}
    \label{tab:isotopes}
\end{table}

The obtained spectra can be used as input to simulate the detector activation due to hadron interactions (\cite{4774966, 1989AIPC..186..278D, 1993AIPC..280.1107B}). The simulations were performed using the Medium Energy Gamma-ray Astronomy library (MEGAlib) toolkit (\cite{2006NewAR..50..629Z}). They are divided in three steps:
\begin{enumerate}
    \item Simulation and storage of the isotopes generated by the passage of the initial particle through the considered spacecraft or material.
    \item Calculation of the activation radiation after a certain irradiation time. The irradiation can either be constant during the whole time, or be followed by a certain cool down time.
    \item Simulation of the isotopes decay after the irradiation.
\end{enumerate}
The spacecraft used in the simulation is a mass model of the gamma-ray mission proposal e-ASTROGAM (\cite{2017ExA....44...25D}).\\
Dedicated to the gamma-ray observation in the 300 keV to 3 GeV energy range, the e-ASTROGAM instrument will consist of three main detectors: a tracker, a calorimeter and an anticoincidence system. The tracker is foreseen to be composed of 56 silicon layers supported on their side by a mechanical structure. The calorimeter is composed by thallium activated cesium iodide crystals, read out by silicon drift detectors. The anticoincidence, covering both the top and the side of the instrument and used to veto the triggers from charged particles, is made by plastic scintillator coupled to silicon photomultipliers. Thanks to this geometry, e-ASTROGAM will be able to reconstruct gamma-ray events in both the Compton and pair regime.\\
\subsection{Simulations}
The simulated particles for the activation simulations were: primary and secondary protons, alpha particles, atmospheric neutrons, and protons trapped in the SAA.\\
For the primary and secondary protons, as well as for the atmospheric neutrons and alpha particles, from which the irradiation can be considered steady along a given orbit, a constant irradiation for 1 year was simulated. The spectrum used was the one described in sec. \ref{s:model}, supposing an equatorial orbit with 550 km of altitude.\\
In the case of the SAA protons, since with MEGAlib it is not yet possible to simulate the variation of the spectrum at the passage of the satellite through the area, the average spectrum of a one month observation, shown in Fig. \ref{fig:SAAflux} was used. The short and long term activation were studied separately in the case of the SAA. For the short term, the one affecting observation at every orbit, just after the passage through the anomaly, simulations were carried out for 10 minutes of constant irradiation with some cool down time (0, 1, 2, 5, 10 minutes). The cool down corresponds to a time after the end of the irradiation during which the data taking has not yet started. Only 10 minutes of the total passage through the SAA ($\sim19$ minutes for a 550 km equatorial orbit) were simulated, since the contribution of the isotopes created at earlier times is, given the short lifetime, negligible. For the long term activation, simulations were performed for a total of $\sim72$ days, equivalent to the cumulative passage through the SAA in one year. A half a orbit cool down was then added to the simulation in order to consider only the long-lived isotopes contribution.\\
Simulations were made for two different altitudes (550 and 600 km) and several orbit inclinations ($0^\circ, 1^\circ, 2^\circ, 3^\circ, 4^\circ, 5^\circ, 10^\circ, 15^\circ$). Tab. \ref{tab:isotopes} collects the most abundant isotopes created by the passage of the particles inside the different parts of the satellite.\\

\subsection{Analysis: count rates}

After the simulation, the data were analyzed in order to retrieve the number of events erroneously tagged as gamma-ray, either in the Compton or pair regime. The rate of the falsely reconstructed pair events is extremely low ($<0.1$ counts/s). This is mainly due to the low energy nature of the isotopes' decay.\\
\begin{figure}[ht]
\begin{center}
\includegraphics[width=0.95\textwidth]{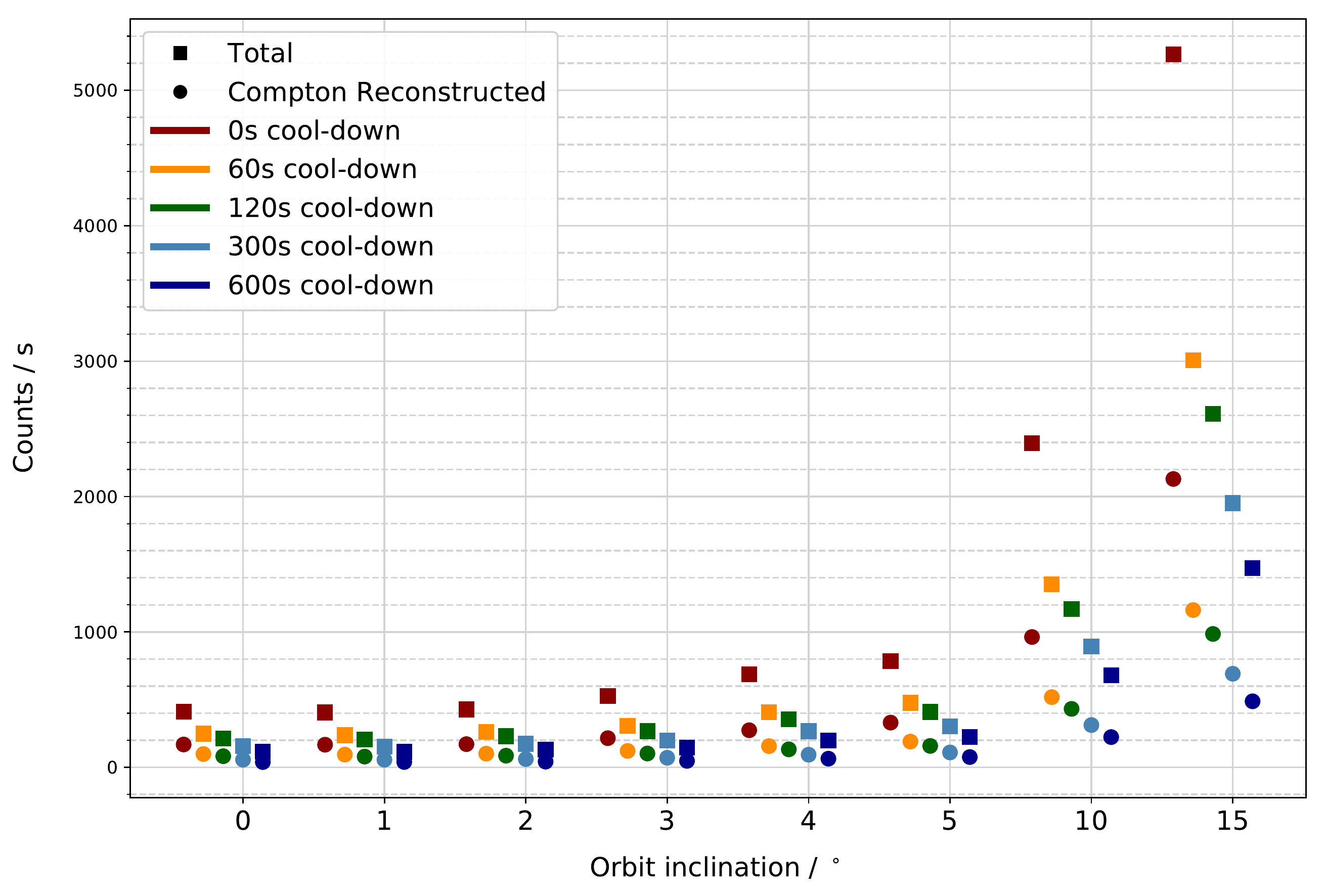}
\caption{Analysis of the short-term activation caused by the passage of the SAA at 550, considering different cool down times and orbit inclinations. For the same inclination and cool down time, the two markers corresponds to the total number of events (squares) and the number of events reconstructed as Compton events (circles).}
\label{fig:allACVeto}
\end{center}
\end{figure}

The total rate of decays of short-lived isotopes and the rate of the Compton reconstructed events related to the passage through the SAA are shown in Fig. \ref{fig:allACVeto}. Small changes are visible for orbits with inclinations smaller than $\sim5^\circ$, with a $<1.5$ ratio between the results at $0^\circ$ and $5^\circ$ inclination. A more than a factor of two increase is instead observed in the reconstructed events at 10$^\circ$ and 15$^\circ$. The increase is consistent with the raise in the observed flux shown in Fig. \ref{fig:SAAflux}. As expected, by considering a non-zero cool down time, the remaining events decreases following an exponential trend.\\ 
\begin{figure}[ht]
\begin{center}
\includegraphics[width=0.95\textwidth]{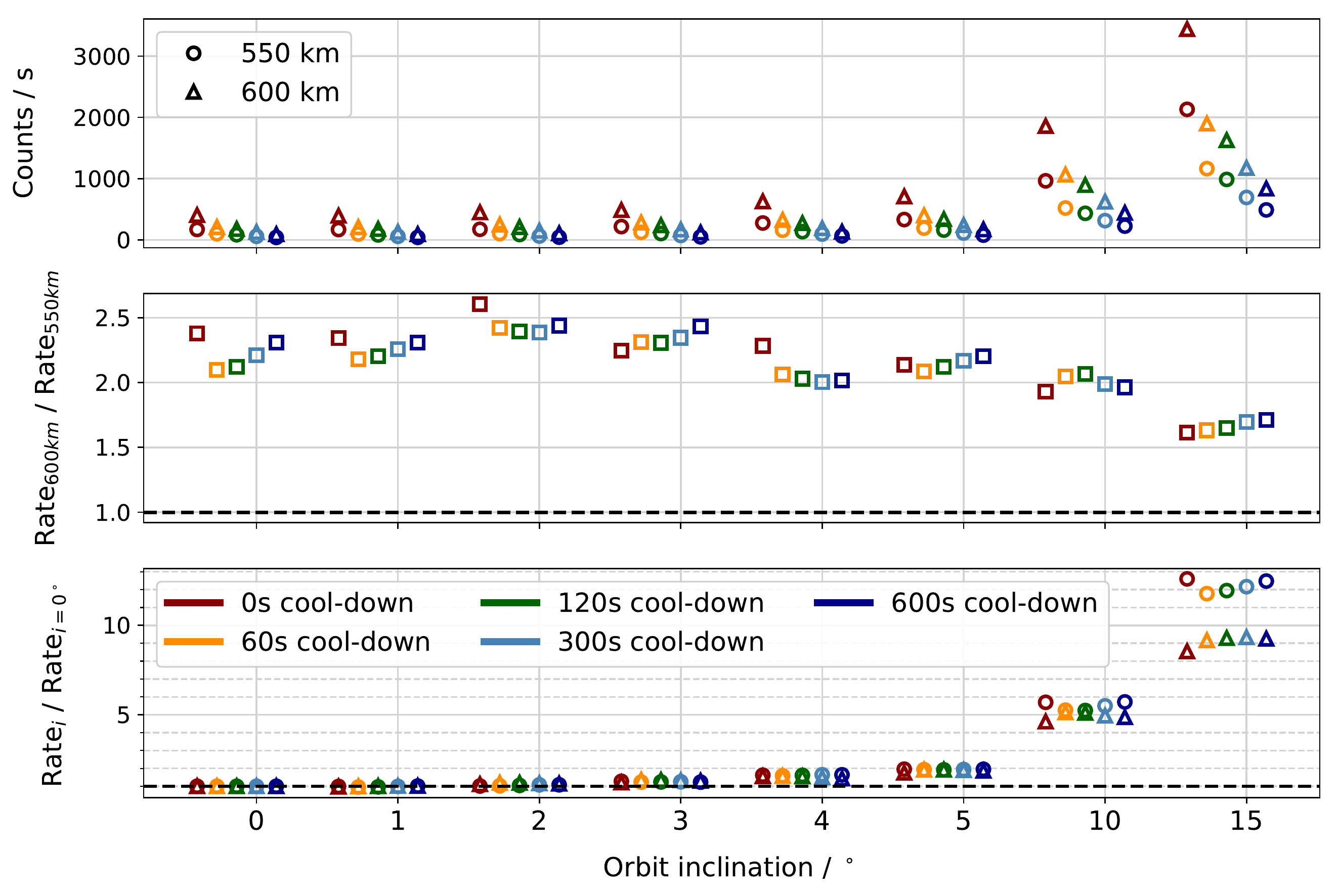}
\caption{Analysis of the activation from short-lived isotopes caused by the passage of the SAA at 550 km (circles) and 600 km (triangles), considering different cool down times and orbit inclinations. The different cool down times are shown in different colors, as in Fig. \ref{fig:allACVeto}. \textit{Top}: Number of events reconstructed as Compton events. \textit{Center}: Ratio of the values obtained at the two altitudes. \textit{Bottom}: Ratio with respect to the perfectly equatorial orbit.}
\label{fig:allRecCom}
\end{center}
\end{figure}
The rates increase with the altitude, as shown in Fig. \ref{fig:allRecCom}, with a factor between 1.5 and 2.5, depending on the inclination of the orbit $i$. The rates increase with the inclination of the orbit $i$, with almost constant results up to $i\sim3^\circ$, gradually rising to a factor of 2 at $5^\circ$, a factor of 5 at $5^\circ$, and reaching an order of magnitude difference for an inclination of $15^\circ$. These changes do not depend on the trigger choice. Either implementing a strict veto on the anticoincidence, i.e. without using its segmentation but vetoing all the events with a signal in any of the panel, or accepting all reconstructed events, the ratio remains constant and the most favourable orbit is always a 550 km orbit with $i \lesssim 4^\circ$. The change in the applied veto affects only the absolute number of reconstructed events, with a $\sim$25\% difference between the two mentioned cases. It does not influence the rates variation at different inclination or altitudes.\\

\begin{figure}[ht]
\begin{center}
\includegraphics[width=0.95\textwidth]{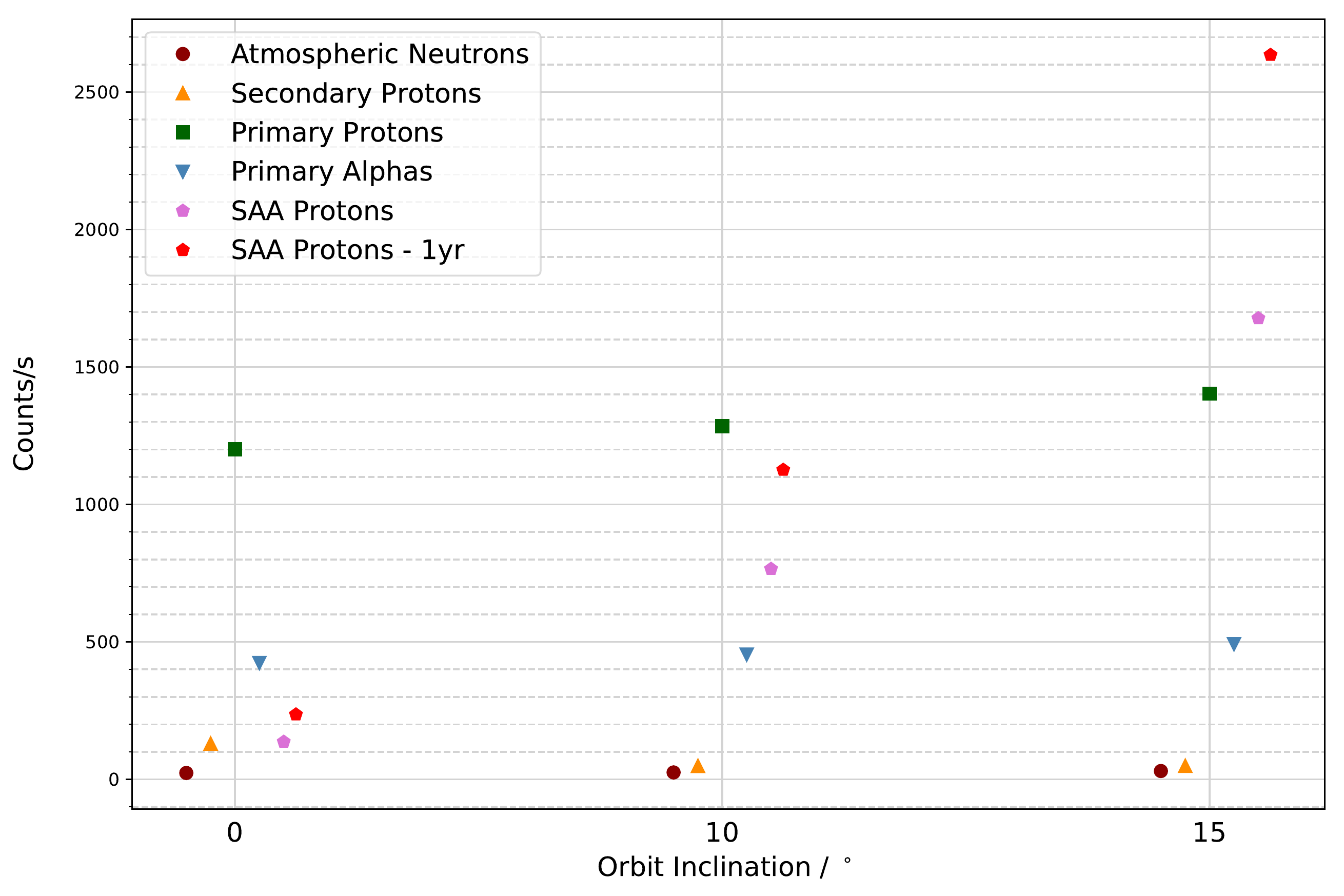}
\caption{Comparison of the results of the activation caused by the general on-orbit background and by the passage of the SAA, both from the short-lived isotopes, without considering any cool down time, and from the long-live ones. Results are shown for three different orbit inclinations. All the results were calculated for an altitude of 550 km.}
\label{fig:ASGB}
\end{center}
\end{figure}

The rate of events reconstructed as Compton for the general background is shown in Fig. \ref{fig:ASGB}, together with the results from one SAA passage without cool down time, and its long-term activation after one year in orbit.\\ 
As shown in Fig. \ref{fig:incaltchange} (bottom panel), the flux of the secondary protons at $i=10^\circ$ is significantly lower than that at $i=0^\circ$ for $E<5$ GeV. It only increases at higher energies, leading to a total factor of 3 decrease in the event rates at $10^\circ$. Between $10^\circ$ and $15^\circ$, no changes in the secondary spectra are observed. The atmospheric neutrons spectrum changes by only $\sim20\%$, causing a $\sim15\%$ increase in the rates. Changing from $i=0^\circ$ to $i=10^\circ$, both the primary protons and alpha particles spectra change by more than a factor of 2 for energies lower than the geomagnetic cutoff. This low energy increase does not affect significantly the remaining counts, which change by only $\sim10\%$. Instead the SAA protons spectrum increases by a factor 2, at low energies, to a factor of $\sim 7$, for energies higher than 10 MeV, which, as previously shown, leads to a factor of 5 difference in the short-term rates.\\ 
Up to an inclination of $10^\circ$, the main contributor to the total activation is the decay of the isotopes created by the primary protons, the secondary protons contribution being an order of magnitude smaller than that. The counts from the alpha particles passage are one third of the ones related to the primary protons, while the atmospheric neutrons rates are negligible.\\
The total SAA contribution, without considering any cool down time, can be considered negligible for low inclinations ($i<5^\circ$), but it becomes more important than the alpha particles contribution for $i\geqslant 10^\circ$, and the main contributor at $15^\circ$. For $i\geqslant 10^\circ$ it is necessary to consider some cool down time for the activation rate from the trapped protons to become negligible. Instead, for low inclination orbit, the short-term SAA activation contribution, comparable to the one from the secondary protons, is always smaller than the contribution of both the primary protons and alpha particles. The long-term SAA activation contribution is higher than the short-term by a factor $\sim1.5-1.6$ in the considered inclination range. The SAA is the main contributor to the background rates for high-inclination orbits, surpassing the rates from the primary protons by a factor of 1.9.\\

\begin{figure}[ht]
\begin{center}
\includegraphics[width=0.935\textwidth]{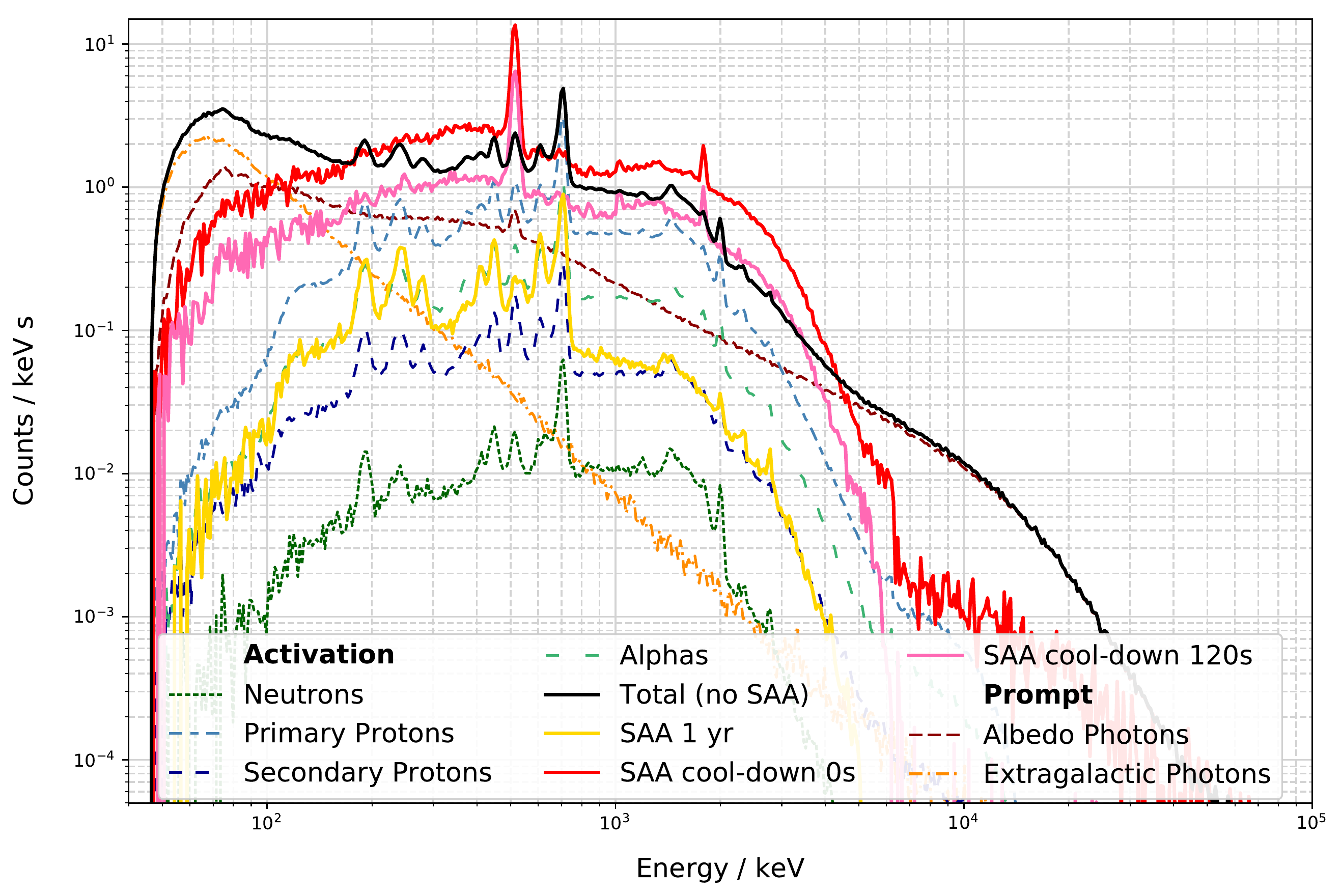}\\
\includegraphics[width=0.935\textwidth]{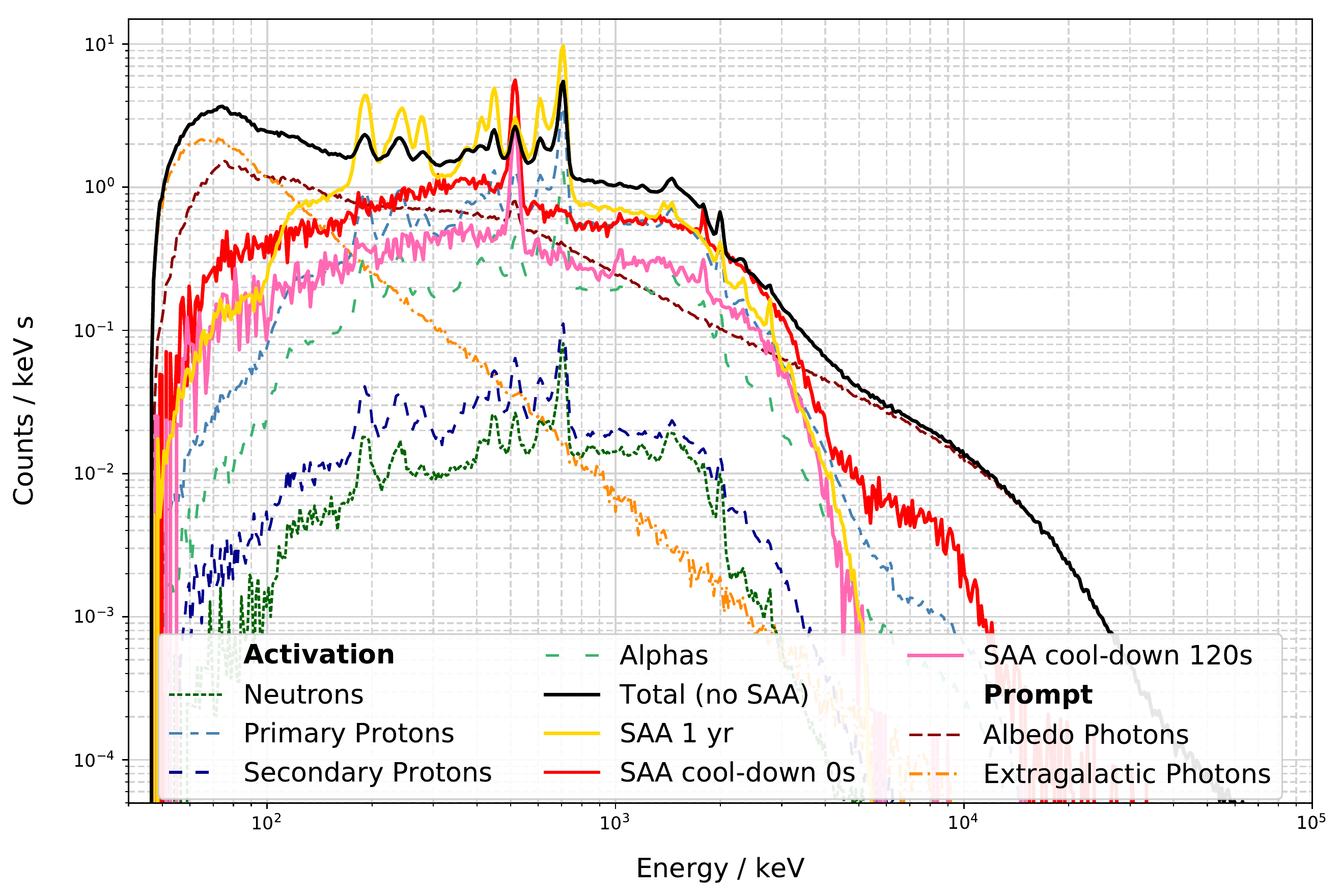}
\caption{Spectra of the reconstructed background Compton events for an orbit inclination of 0$^\circ$ (\textit{top}) or 15$^\circ$ (\textit{bottom}). The total is calculated as the sum of all the component from the activation except the SAA contribution. The SAA contribution is divided in short-term, with a 0 s or 120 s cool down, and long-term (SAA 1 yr).}
\label{fig:spectra}
\end{center}
\end{figure}
\subsection{Analysis: spectra}

The spectra of the reconstructed Compton events with and without a recoil electron track, from the activation simulations and from the extragalactic and albedo photons along an orbit inclined by 0$^\circ$ and 15$^\circ$ are shown in Fig. \ref{fig:spectra}. A strict veto is used for both the side and top anticoincidence but no quality cuts are applied to the reconstructed events.\\
The albedo and extragalactic photons are the main components of the background until $\sim 400$~keV, afterward passed by the primary protons until $\sim~3$~MeV. The extragalactic photons contribution, while being the main one below 100~keV, it quickly loses importance at higher energies.\\
While the total rate of the events related to the passage through the SAA is negligible for an equatorial orbit, they must be considered in the estimation of the background for the 511 keV line observation. The contribution to this line background from the short-term activation in an equatorial orbit represents a fifth of the summed contributions of the other components without cool down, and a tenth after two minutes. At an inclination of $15^\circ$ the short-term SAA contribution is not only dominant when no cool down is considered, five times the sum of the other components, but it is still comparable to the total after a two minutes cool-down time. The long-term activation as well contributes to the 511 keV line background, similarly to the short-term contribution when considering a two minutes cool down. For an inclination of 15$^\circ$, the long-term contribution to the background is higher than the sum of all the other components in the $\sim150$ keV to $\sim720$ keV energy range. The higher inclination orbits are therefore disfavoured.\\
The two most prominent lines that appear in the spectra are the 511~keV and 700~keV lines. The 511~keV is created by positrons annihilation. Positrons are the results of the $\beta ^+$ decay of isotopes such as $^{11}$C or $^{15}$O. The 700 keV line is visible in all activation spectra except in the SAA short-term ones. It is mostly generated by the electron-capture decay of  two long-lived isotopes abundantly created in the CsI(Tl) crystals of the calorimeter: $^{132}$Cs and $^{126}$I (see Table \ref{tab:isotopes}). The first radioisotope produces nuclear gamma rays of 668 keV at the same time as K$\alpha$ X-rays of 30 keV. The second one produces 666 keV gamma rays together with 27 keV X-rays.

\section{Conclusions}\label{s:Con}
Thanks to the shielding provided by the geomagnetic field, Low Earth Orbits (LEOs) present the lowest possible background for a gamma-ray space mission. Satellites on LEO are nonetheless affected by a particle flux which depends on both the altitude and the inclination of the orbit. This background has been modeled starting from data of present and past experiments, considering the possibility of small changes in the orbit altitude and inclination. The model was later used to estimate the count rate from the decay of isotopes created by the passage of these particles inside a possible future gamma-ray mission, e-ASTROGAM (\cite{2017ExA....44...25D}), using the tool MEGAlib (\cite{2006NewAR..50..629Z}). This event rate was compared to the one from the residual activity after a passage through the South Atlantic Anomaly (SAA), and its cumulative effect after one year. The spectrum of the SAA has been computed using the AE9/AP9-IRENE (\cite{Ginet2013}) model.\\
The decay of the isotopes created by the primary cosmic-ray protons is found to be the main contributor of the activation, the rates from alpha particles being a fourth of that, while the contributions of the neutrons and secondary protons can be considered negligible. The SAA contribution gains importance when increasing the inclination of the orbit: from being negligible in nearly equatorial orbit, it becomes the second contributor for a $10^\circ$ inclination, surpassing the event rate from the alpha particles interaction, and the main contributor at $15^\circ$. \\
While the total event rate might be negligible, the SAA background for the 511 keV line observation must be always taken in consideration. While at low inclination the short-term contribution to the line becomes quickly negligible, at high inclination a long cool down is needed for the rates to drop. For a $15^\circ$ inclination, the long-term SAA contribution is also higher than the sum of all the other components in the 150 to 720 keV energy range, while it is comparable to the alpha particles contribution for an equatorial orbit.\\ 
A lower background and activation rate is found for an altitude of 550 km than for 600 km, regardless of the inclination. The changes for a 50 km increase in altitude are small in the case of the general background, but a factor between two and three in the SAA case.\\
The best LEO for a gamma-ray mission is therefore found to be a low inclination ($i<5$) and low altitude (550 km) orbit. Although, it should be considered that a study on the atmospheric drag at the different altitudes, while important in the planning of any LEO mission, is beyond the scope of this paper. This kind of study might change the preference for a lower altitude orbit but should not affect the obtained result on its inclination.

\begin{acknowledgements}
This work has been carried out in the framework of the H2020 project AHEAD, funded by the European Union.
\end{acknowledgements}

\bibliography{BackgroundLEO}

\newcommand{\noop}[1]{}
\begin{thebibliography}{30}
\providecommand{\natexlab}[1]{#1}
\providecommand{\url}[1]{{#1}}
\providecommand{\urlprefix}{URL }
\expandafter\ifx\csname urlstyle\endcsname\relax
  \providecommand{\doi}[1]{DOI~\discretionary{}{}{}#1}\else
  \providecommand{\doi}{DOI~\discretionary{}{}{}\begingroup
  \urlstyle{rm}\Url}\fi
\providecommand{\eprint}[2][]{\url{#2}}

\bibitem[{{Abdo} et~al(2009{\natexlab{a}}){Abdo}, {Ackermann}, {Ajello},
  {Ampe}, {Anderson}, {Atwood}, {Axelsson}, {Bagagli}, {Baldini}, {Ballet}, and
  et~al.}]{2009APh....32..193A}
{Abdo} AA, {Ackermann} M, {Ajello} M, {Ampe} J, {Anderson} B, {Atwood} WB,
  {Axelsson} M, {Bagagli} R, {Baldini} L, {Ballet} J, et~al
  (2009{\natexlab{a}}) {The on-orbit calibration of the Fermi Large Area
  Telescope}. Astroparticle Physics 32:193--219,
  \doi{10.1016/j.astropartphys.2009.08.002}, \eprint{arXiv: 0904.2226}

\bibitem[{{Abdo} et~al(2009{\natexlab{b}}){Abdo}, {Ackermann}, {Ajello},
  {Atwood}, {Baldini}, {Ballet}, {Barbiellini}, {Bastieri}, {Baughman},
  {Bechtol}, and et~al.}]{2009PhRvD..80l2004A}
{Abdo} AA, {Ackermann} M, {Ajello} M, {Atwood} WB, {Baldini} L, {Ballet} J,
  {Barbiellini} G, {Bastieri} D, {Baughman} BM, {Bechtol} K, et~al
  (2009{\natexlab{b}}) {Fermi large area telescope observations of the
  cosmic-ray induced {$\gamma$}-ray emission of the Earth's atmosphere}. \prd,
  80(12):122004, \doi{10.1103/PhysRevD.80.122004}, \eprint{arXiv: 0912.1868}

\bibitem[{{Ackermann} et~al(2014){Ackermann}, {Ajello}, {Albert}, {Allafort},
  {Baldini}, {Barbiellini}, {Bastieri}, {Bechtol}, {Bellazzini}, {Blandford},
  and {Fermi LAT Collaboration}}]{2014PhRvL.112o1103A}
{Ackermann} M, {Ajello} M, {Albert} A, {Allafort} A, {Baldini} L, {Barbiellini}
  G, {Bastieri} D, {Bechtol} K, {Bellazzini} R, {Blandford} RD, {Fermi LAT
  Collaboration} (2014) {Inferred Cosmic-Ray Spectrum from Fermi Large Area
  Telescope {$\gamma$}-Ray Observations of Earth's Limb}. Physical Review
  Letters 112(15):151103, \doi{10.1103/PhysRevLett.112.151103}, \eprint{arXiv:
  1403.5372}

\bibitem[{{Ackermann} et~al(2015){Ackermann}, {Ajello}, {Albert}, {Atwood},
  {Baldini}, {Ballet}, {Barbiellini}, {Bastieri}, {Bechtol}, {Bellazzini},
  {Bissaldi}, and et~al.}]{2015ApJ...799...86A}
{Ackermann} M, {Ajello} M, {Albert} A, {Atwood} WB, {Baldini} L, {Ballet} J,
  {Barbiellini} G, {Bastieri} D, {Bechtol} K, {Bellazzini} R, {Bissaldi} E,
  et~al (2015) {The Spectrum of Isotropic Diffuse Gamma-Ray Emission between
  100 MeV and 820 GeV}. \apj, 799:86, \doi{10.1088/0004-637X/799/1/86},
  \eprint{arXiv: 1410.3696}

\bibitem[{Aguilar et~al(2014)Aguilar, Aisa, Alvino, Ambrosi, Andeen, Arruda,
  Attig, Azzarello, Bachlechner, Barao, and et~al.}]{PhysRevLett.113.121102}
Aguilar M, Aisa D, Alvino A, Ambrosi G, Andeen K, Arruda L, Attig N, Azzarello
  P, Bachlechner A, Barao F, et~al (2014) Electron and positron fluxes in
  primary cosmic rays measured with the alpha magnetic spectrometer on the
  international space station. Phys Rev Lett 113:121102,
  \doi{10.1103/PhysRevLett.113.121102},
  \urlprefix\url{https://link.aps.org/doi/10.1103/PhysRevLett.113.121102}

\bibitem[{Aguilar et~al(2015{\natexlab{a}})Aguilar, Aisa, Alpat, Alvino,
  Ambrosi, Andeen, Arruda, Attig, Azzarello, Bachlechner, and
  et~al.}]{PhysRevLett.115.211101}
Aguilar M, Aisa D, Alpat B, Alvino A, Ambrosi G, Andeen K, Arruda L, Attig N,
  Azzarello P, Bachlechner A, et~al (2015{\natexlab{a}}) Precision measurement
  of the helium flux in primary cosmic rays of rigidities 1.9 gv to 3 tv with
  the alpha magnetic spectrometer on the international space station. Phys Rev
  Lett 115:211101, \doi{10.1103/PhysRevLett.115.211101},
  \urlprefix\url{https://link.aps.org/doi/10.1103/PhysRevLett.115.211101}

\bibitem[{Aguilar et~al(2015{\natexlab{b}})Aguilar, Aisa, Alpat, Alvino,
  Ambrosi, Andeen, Arruda, Attig, Azzarello, Bachlechner, and
  et~al.}]{PhysRevLett.114.171103}
Aguilar M, Aisa D, Alpat B, Alvino A, Ambrosi G, Andeen K, Arruda L, Attig N,
  Azzarello P, Bachlechner A, et~al (2015{\natexlab{b}}) Precision measurement
  of the proton flux in primary cosmic rays from rigidity 1 gv to 1.8 tv with
  the alpha magnetic spectrometer on the international space station. Phys Rev
  Lett 114:171103, \doi{10.1103/PhysRevLett.114.171103},
  \urlprefix\url{https://link.aps.org/doi/10.1103/PhysRevLett.114.171103}

\bibitem[{Alcaraz et~al(2000{\natexlab{a}})Alcaraz, Alpat, Ambrosi, Anderhub,
  Ao, Arefiev, Azzarello, Babucci, Baldini, Basile, and et~al.}]{200010}
Alcaraz J, Alpat B, Ambrosi G, Anderhub H, Ao L, Arefiev A, Azzarello P,
  Babucci E, Baldini L, Basile M, et~al (2000{\natexlab{a}}) Leptons in near
  earth orbit. Physics Letters B 484(1):10 -- 22,
  \doi{https://doi.org/10.1016/S0370-2693(00)00588-8},
  \urlprefix\url{http://www.sciencedirect.com/science/article/pii/S0370269300005888}

\bibitem[{Alcaraz et~al(2000{\natexlab{b}})Alcaraz, Alvisi, Alpat, Ambrosi,
  Anderhub, Ao, Arefiev, Azzarello, Babucci, Baldini, and et~al.}]{2000215}
Alcaraz J, Alvisi D, Alpat B, Ambrosi G, Anderhub H, Ao L, Arefiev A, Azzarello
  P, Babucci E, Baldini L, et~al (2000{\natexlab{b}}) Protons in near earth
  orbit. Physics Letters B 472(1):215 -- 226,
  \doi{https://doi.org/10.1016/S0370-2693(99)01427-6},
  \urlprefix\url{http://www.sciencedirect.com/science/article/pii/S0370269399014276}

\bibitem[{Armstrong et~al(1973)Armstrong, Chandler, and
  Barish}]{JA078i016p02715}
Armstrong TW, Chandler KC, Barish J (1973) Calculations of neutron flux spectra
  induced in the earth's atmosphere by galactic cosmic rays. Journal of
  Geophysical Research 78(16):2715--2726, \doi{10.1029/JA078i016p02715},
  \urlprefix\url{https://agupubs.onlinelibrary.wiley.com/doi/abs/10.1029/JA078i016p02715},
  \eprint{https://agupubs.onlinelibrary.wiley.com/doi/pdf/10.1029/JA078i016p02715}

\bibitem[{{Atwood} et~al(2009){Atwood}, {Abdo}, {Ackermann}, {Althouse},
  {Anderson}, {Axelsson}, {Baldini}, {Ballet}, {Band}, {Barbiellini}, and
  et~al.}]{2009ApJ...697.1071A}
{Atwood} WB, {Abdo} AA, {Ackermann} M, {Althouse} W, {Anderson} B, {Axelsson}
  M, {Baldini} L, {Ballet} J, {Band} DL, {Barbiellini} G, et~al (2009) {The
  Large Area Telescope on the Fermi Gamma-Ray Space Telescope Mission}. \apj
  697:1071--1102, \doi{10.1088/0004-637X/697/2/1071}, \eprint{arXiv: 0902.1089}

\bibitem[{{Battersby} et~al(1993){Battersby}, {Quenby}, {Dyer}, {Truscott},
  {Hammond}, {Comber}, {Kurfess}, {Johnson}, {Kinzer}, {Strickman}, {Jung},
  {Purcell}, {Grabelsky}, and {Ulmer}}]{1993AIPC..280.1107B}
{Battersby} SJR, {Quenby} JJ, {Dyer} CS, {Truscott} PR, {Hammond} NDA, {Comber}
  C, {Kurfess} JD, {Johnson} WN, {Kinzer} RL, {Strickman} MS, {Jung} GV,
  {Purcell} WR, {Grabelsky} DA, {Ulmer} MP (1993) {Calculation of the induced
  radioactivity background in OSSE.} In: {Friedlander} M, {Gehrels} N, {Macomb}
  DJ (eds) American Institute of Physics Conference Series, American Institute
  of Physics Conference Series, vol 280, pp 1107--1111, \doi{10.1063/1.44183}

\bibitem[{{Churazov} et~al(2006){Churazov}, {Sazonov}, {Sunyaev}, and
  {Revnivtsev}}]{2006astro.ph..8252C}
{Churazov} E, {Sazonov} S, {Sunyaev} R, {Revnivtsev} M (2006) {Earth X-ray
  albedo for cosmic X-ray background radiation in the 1--1000 keV band}. ArXiv
  Astrophysics e-prints \eprint{astro-ph/0608252}

\bibitem[{{De Angelis} et~al(2017){De Angelis}, {Tatischeff}, {Tavani},
  {Oberlack}, {Grenier}, {Hanlon}, {Walter}, {Argan}, {von Ballmoos},
  {Bulgarelli}, and et~al.}]{2017ExA....44...25D}
{De Angelis} A, {Tatischeff} V, {Tavani} M, {Oberlack} U, {Grenier} I, {Hanlon}
  L, {Walter} R, {Argan} A, {von Ballmoos} P, {Bulgarelli} A, et~al (2017) {The
  e-ASTROGAM mission. Exploring the extreme Universe with gamma rays in the MeV
  - GeV range}. Experimental Astronomy 44:25--82,
  \doi{10.1007/s10686-017-9533-6}, \eprint{arXiv: 1611.02232}

\bibitem[{{De Angelis} et~al(2018){De Angelis}, {Tatischeff}, {Grenier},
  {McEnery}, {Mallamaci}, {Tavani}, {Oberlack}, {Hanlon}, {Walter}, {Argan},
  and et~al.}]{2017arXiv171101265D}
{De Angelis} A, {Tatischeff} V, {Grenier} IA, {McEnery} J, {Mallamaci} M,
  {Tavani} M, {Oberlack} U, {Hanlon} L, {Walter} R, {Argan} A, et~al (2018)
  {Science with e-ASTROGAM (A space mission for MeV-GeV gamma-ray
  astrophysics)}. To be published in Journal of High Energy Astrophysics
  \eprint{arXiv: 1711.01265}

\bibitem[{{Dyer} et~al(1989){Dyer}, {Truscott}, {Hammond}, and
  {Comber}}]{1989AIPC..186..278D}
{Dyer} CS, {Truscott} PR, {Hammond} NDA, {Comber} C (1989) {Radioactivity
  induced in gamma-ray spectrometers}. In: {Rester} AC Jr, {Trombka} JI (eds)
  High-Energy Radiation Background in Space, American Institute of Physics
  Conference Series, vol 186, pp 278--288, \doi{10.1063/1.38187}

\bibitem[{Ginet et~al(2013)Ginet, O'Brien, Huston, Johnston, Guild, Friedel,
  Lindstrom, Roth, Whelan, Quinn, and et~al.}]{Ginet2013}
Ginet GP, O'Brien TP, Huston SL, Johnston WR, Guild TB, Friedel R, Lindstrom
  CD, Roth CJ, Whelan P, Quinn RA, et~al (2013) Ae9, ap9 and spm: New models
  for specifying the trapped energetic particle and space plasma environment.
  Space Science Reviews 179(1):579--615, \doi{10.1007/s11214-013-9964-y},
  \urlprefix\url{https://doi.org/10.1007/s11214-013-9964-y}

\bibitem[{{Kole} et~al(2015){Kole}, {Pearce}, and {Mu{\~n}oz
  Salinas}}]{2015APh....62..230K}
{Kole} M, {Pearce} M, {Mu{\~n}oz Salinas} M (2015) {A model of the cosmic ray
  induced atmospheric neutron environment}. Astroparticle Physics 62:230--240,
  \doi{10.1016/j.astropartphys.2014.10.002}, \eprint{arXiv: 1410.1364}

\bibitem[{Lingenfelter(1963)}]{JZ068i020p05633}
Lingenfelter RE (1963) The cosmic-ray neutron leakage flux. Journal of
  Geophysical Research 68(20):5633--5639, \doi{10.1029/JZ068i020p05633},
  \urlprefix\url{https://agupubs.onlinelibrary.wiley.com/doi/abs/10.1029/JZ068i020p05633},
  \eprint{https://agupubs.onlinelibrary.wiley.com/doi/pdf/10.1029/JZ068i020p05633}

\bibitem[{Mizuno et~al(2004)Mizuno, Kamae, Godfrey, Handa, Thompson, Lauben,
  Fukazawa, and Ozaki}]{0004-637X-614-2-1113}
Mizuno T, Kamae T, Godfrey G, Handa T, Thompson DJ, Lauben D, Fukazawa Y, Ozaki
  M (2004) Cosmic-ray background flux model based on a gamma-ray large area
  space telescope balloon flight engineering model. The Astrophysical Journal
  614(2):1113, \urlprefix\url{http://stacks.iop.org/0004-637X/614/i=2/a=1113}

\bibitem[{O'Brien et~al(2018)O'Brien, Johnston, Huston, Roth, Guild, Su, and
  Quinn}]{8101531}
O'Brien TP, Johnston WR, Huston SL, Roth CJ, Guild TB, Su YJ, Quinn RA (2018)
  Changes in ae9/ap9-irene version 1.5. IEEE Transactions on Nuclear Science
  65(1):462--466, \doi{10.1109/TNS.2017.2771324}

\bibitem[{{Sazonov} et~al(2007){Sazonov}, {Churazov}, {Sunyaev}, and
  {Revnivtsev}}]{2007MNRAS.377.1726S}
{Sazonov} S, {Churazov} E, {Sunyaev} R, {Revnivtsev} M (2007) {Hard X-ray
  emission of the Earth's atmosphere: Monte Carlo simulations}. \mnras
  377:1726--1736, \doi{10.1111/j.1365-2966.2007.11746.x},
  \eprint{astro-ph/0608253}

\bibitem[{Smart and Shea(2005)}]{SMART20052012}
Smart D, Shea M (2005) A review of geomagnetic cutoff rigidities for
  earth-orbiting spacecraft. Advances in Space Research 36(10):2012 -- 2020,
  \doi{https://doi.org/10.1016/j.asr.2004.09.015},
  \urlprefix\url{http://www.sciencedirect.com/science/article/pii/S0273117705001997},
  solar Wind-Magnetosphere-Ionosphere Dynamics and Radiation Models

\bibitem[{Sreekumar et~al(1998)Sreekumar, Bertsch, Dingus, Esposito, Fichtel,
  Hartman, Hunter, Kanbach, Kniffen, Lin, Mayer-Hasselwander, Michelson, von
  Montigny, Mücke, Mukherjee, Nolan, Pohl, Reimer, Schneid, Stacy, Stecker,
  Thompson, and Willis}]{0004-637X-494-2-523}
Sreekumar P, Bertsch DL, Dingus BL, Esposito JA, Fichtel CE, Hartman RC, Hunter
  SD, Kanbach G, Kniffen DA, Lin YC, Mayer-Hasselwander HA, Michelson PF, von
  Montigny C, Mücke A, Mukherjee R, Nolan PL, Pohl M, Reimer O, Schneid E,
  Stacy JG, Stecker FW, Thompson DJ, Willis TD (1998) Egret observations of the
  extragalactic gamma-ray emission. The Astrophysical Journal 494(2):523,
  \urlprefix\url{http://stacks.iop.org/0004-637X/494/i=2/a=523}

\bibitem[{{Thompson} et~al(1981){Thompson}, {Simpson}, and
  {Ozel}}]{1981JGR....86.1265T}
{Thompson} DJ, {Simpson} GA, {Ozel} ME (1981) {SAS 2 observations of the earth
  albedo gamma radiation above 35 MeV}. \jgr 86:1265--1270,
  \doi{10.1029/JA086iA03p01265}

\bibitem[{{T{\"u}rler} et~al(2010){T{\"u}rler}, {Chernyakova}, {Courvoisier},
  {Lubi{\'n}ski}, {Neronov}, {Produit}, and {Walter}}]{2010A&A...512A..49T}
{T{\"u}rler} M, {Chernyakova} M, {Courvoisier} TJL, {Lubi{\'n}ski} P, {Neronov}
  A, {Produit} N, {Walter} R (2010) {INTEGRAL hard X-ray spectra of the cosmic
  X-ray background and Galactic ridge emission}. \aap, 512:A49,
  \doi{10.1051/0004-6361/200913072}, \eprint{arXiv: 1001.2110}

\bibitem[{{van Allen}(1958)}]{allen1958observation}
{van Allen} JA (1958) Observation of high intensity radiation by satellites
  1958 alpha and gamma. Journal of Jet Propulsion 28(9):588--592

\bibitem[{{Winkler} et~al(2003){Winkler}, {Courvoisier}, {Di Cocco}, {Gehrels},
  {Gim{\'e}nez}, {Grebenev}, {Hermsen}, {Mas-Hesse}, {Lebrun}, {Lund}, and
  et~al.}]{2003A&A...411L...1W}
{Winkler} C, {Courvoisier} TJL, {Di Cocco} G, {Gehrels} N, {Gim{\'e}nez} A,
  {Grebenev} S, {Hermsen} W, {Mas-Hesse} JM, {Lebrun} F, {Lund} N, et~al (2003)
  {The INTEGRAL mission}. \aap 411:L1--L6, \doi{10.1051/0004-6361:20031288}

\bibitem[{{Zoglauer} et~al(2006){Zoglauer}, {Andritschke}, and
  {Schopper}}]{2006NewAR..50..629Z}
{Zoglauer} A, {Andritschke} R, {Schopper} F (2006) {MEGAlib The Medium Energy
  Gamma-ray Astronomy Library}. \nar 50:629--632,
  \doi{10.1016/j.newar.2006.06.049}

\bibitem[{Zoglauer et~al(2008)Zoglauer, Weidenspointner, Wunderer, and
  Boggs}]{4774966}
Zoglauer A, Weidenspointner G, Wunderer CB, Boggs SE (2008) Status of
  instrumental background simulations for gamma-ray telescopes with {G}eant4.
  In: 2008 IEEE Nuclear Science Symposium Conference Record, pp 2859--2864,
  \doi{10.1109/NSSMIC.2008.4774966}

\end{thebibliography}

\end{document}